\documentclass{main}
\title{Graphene-Driven Correlated Electronic States in One Dimensional Defects within WS$_2$}

\author[1,2,3*$\dagger$]{Antonio Rossi}
\author[1,3*$\dagger$]{John C. Thomas}
\author[1,4]{Johannes T. K\"uchle}
\author[1]{Elyse Barr\'{e}}
\author[5,6]{Zhuohang Yu}
\author[7]{Da Zhou}
\author[5,6]{Shalini Kumari}
\author[8]{Hsin-Zon Tsai}
\author[1]{Ed Wong}
\author[2]{Chris Jozwiak}
\author[2]{Aaron Bostwick}
\author[5,6,7,9]{Joshua A. Robinson}
\author[5,6,7,9]{Mauricio Terrones}
\author[1,10]{Archana Raja}
\author[1]{Adam Schwartzberg}
\author[1]{D. Frank Ogletree}
\author[3,8,10]{Jeffrey B. Neaton}
\author[3,8,10]{Michael F. Crommie}
\author[4]{Francesco Allegretti}
\author[4]{Willi Auw\"arter}
\author[2]{Eli Rotenberg} 
\author[1,3*]{Alexander Weber-Bargioni}
\affil[1]{Molecular Foundry, Lawrence Berkeley National Laboratory, Berkeley, CA, 94720, United States of America}
\affil[2]{Advanced Light Source, Lawrence Berkeley National Laboratory, Berkeley, CA, 94720, United States of America}
\affil[3]{Materials Sciences Division, Lawrence Berkeley National Laboratory, Berkeley, CA, 94720, United States of America}
\affil[4]{Physics Department E20, TUM School of Natural Sciences, Technical University of Munich, D-85748, Garching, Germany}
\affil[5]{Department of Materials Science and Engineering, The Pennsylvania State University, University Park, PA, 16082 United States of America}
\affil[6]{Center for Two-Dimensional and Layered Materials, The Pennsylvania State University, University Park, PA, 16802 United States of America}
\affil[7]{Department of Physics, The Pennsylvania State University, University Park, PA, 16802 United States of America}
\affil[8]{Department of Physics, University of California at Berkeley, Berkeley, CA, 94720, United States of America}
\affil[9]{Department of Chemistry, The Pennsylvania State University, University Park, PA, 16802 United States of America}
\affil[10]{Kavli Energy NanoSciences Institute, University of California Berkeley, Berkeley, CA, 94720, United States of America}
\affil[*]{jthomas@lbl.gov, antonio.rossi@iit.it, afweber-bargioni@lbl.gov}
\affil[$\dagger$]{These authors contributed equally.}

\date{}
\begin{document}
\maketitle
\setstcolor{red}
\section*{Abstract}
Tomonaga-Luttinger liquid (TLL) behavior in one-dimensional systems has been predicted and shown to occur at semiconductor-to-metal transitions within two-dimensional materials. Reports of one-dimensional defects hosting a Fermi liquid or a TLL have suggested a dependence on the underlying substrate, however, unveiling the physical details of electronic contributions from the substrate require cross-correlative investigation. Here, we study TLL formation within defectively engineered WS$_2$ atop graphene, where band structure and the atomic environment is visualized with nano angle-resolved photoelectron spectroscopy, scanning tunneling microscopy and spectroscopy, and non-contact atomic force microscopy. Correlations between the local density of states and electronic band dispersion elucidated the electron transfer from graphene into a TLL hosted by one-dimensional metal (1DM) defects. It appears that the vertical heterostructure with graphene and the induced charge transfer from graphene into the 1DM is critical for the formation of a TLL.
\section*{Introduction}

One-dimensional (1D) systems in condensed matter physics provide unique insight into a variety of quasiparticle excitations, including charge density waves (CDWs) that arise due to Peierls instabilities\cite{10.1021/acsnano.0c02072,peierls1955quantum}, lossless transport through electronic wires in topological edge states\cite{Schindler2018,10.1073/pnas.1605982113,Yang2020}, quantum spin liquids\cite{Kim2006}, as well as more exotic phenomena such as Marjorana modes in nanowires\cite{10.1126/science.1222360} and the emergence of a Tomonaga-Luttinger liquid (TLL). The latter has been realized in both nanotubes and transition metal dichalcogenides (TMDs)\cite{10.1143/ptp/5.4.544,10.1063/1.1704046,Bockrath1999,Ishii2003,PhysRevLett.93.166403,10.1126/science.1066266,10.1126/science.1107821,10.1126/science.1171769,PhysRevX.9.011055,10.1021/acsnano.0c05397,Zhu2022,Jia2022} . These quasi particle excitations not only host new condensed matter physics phenomena but hold the promise to become major pillars of quantum electronics and quantum information applications\cite{Keimer2017}. We show the capability to controllably create 1D confined systems and to directly observe TLL formation at its native length scale, which is key to understanding the governing principles behind such a strongly-correlated system.

The hallmarks of a TLL are the independent dispersion of charge and spin, fractional charge transport, and power law suppression of the density of states (DOS) near the Fermi energy (E$_F$)\cite{HASHISAKA201832,PhysRevX.9.011055,10.1021/acsnano.0c05397,Zhu2022}. The formation of a TLL was first observed in 1D carbon nanotubes\cite{Bockrath1999,Ishii2003}, where the conductance of bundled single-walled carbon nanotubes showed power law scaling with respect to bias voltage and temperature. This has since been extended to a number of 1D systems such as semiconductor single-channel wires, nanowires, organic conductors, and fractional quantum Hall edge channels\cite{HASHISAKA201832,10.1126/science.1171769,Kim2006,Bockrath1999,PhysRevLett.113.066105,Ma2017,10.1021/acsnano.5b00410,PhysRevB.100.235403,10.1021/acsnano.7b02172,10.1021/acsnano.0c05397,Du2022}. Recently, this phenomenon has also been shown to exist within 1D defects in two-dimensional (2D) TMDs at a plane of lattice points where the crystal structures on either side of the interface are mirrored, which defines a mirror twin boundary (MTB) within monolayer (ML) TMDs\cite{PhysRevX.9.011055,10.1021/acsnano.0c05397,Zhu2022,Wang2022,10.1021/acs.jpcc.0c01468,10.1021/acs.jpcc.7b08398,10.1021/acsnano.5b00554}. MTB formations have been predicted to exhibit metallic properties and to form out of sub-stoichiometric metal (M = Mo, W) or from depleted chalcogen (X = S, Se, Te) in MX$_2$ materials\cite{10.1021/acs.nanolett.5b02834,Batzill_2018,10.1021/acsnano.5b00410,Barja2016,10.1002/aelm.201600468}. 1D chalcogen vacancy lines and Mo chains have also been shown to exhibit metallic character, where these defects have, in addition, been shown to form into or border other types of 1D defects through mass transport under thermal annealing conditions\cite{Wang2022,10.1021/acsnano.5b00554,10.1021/acsnano.8b01610,HAN2015470,Fang2019,10.1021/acs.jpcc.7b08398}. Each of these defects exhibit metallic behavior, and we now refer to these defects as 1D metals (1DMs), in an otherwise semiconducting TMD material. TLL formation in a 1DM within TMDs adds to the fascinating and vast array of material properties that 2D ML TMDs hold such as single photon emission, tunable band gaps, and strong spin-orbit coupling, to name a few\cite{Lin_2016,10.1146/annurev-physchem-050317-021353,Shimazaki2020,Wang2020,Manzeli2017-nr,Li2018-wt,Ugeda2018,Schuler2020-fh,PhysRevLett.123.076801,Mitterreiter2020-wt,Mitterreiter2021-ih}. Zhu et al. have applied scanning tunneling microscopy and scanning tunneling spectroscopy (STM/STS) to directly map out the local density of states (LDOS) associated with TLLs\cite{Zhu2022}, where measurements show a gap opening near E$_F$, a length dependence on the highest occupied and lowest unoccupied state (HOS/LUS) band gap, and spin-charge dispersion observable in Fourier transform (FT) STS maps\cite{PhysRevX.9.011055,10.1021/acsnano.0c05397}.

The influence of the substrate on TLL formation in 1DMs within a TMD is not yet fully understood nor has it been systematically studied. Reports have shown that 1DM structures placed on Au and graphite show no evidence of TLL \cite{10.1021/acsnano.0c05397}. The hypothesis investigated in this report is that the vertical and direct contact of the 1DM on graphene is critical for the formation of a TLL. In this regard, a clear explanation is needed to connect the macroscopic band structure with the local electronic structure. In order to address these questions, we engineer 1DMs into 2D WS$_2$ grown on graphene with tunable control over both length and density via a post-synthesis, in-situ approach. Measurement techniques cross-correlating STM/STS together with spatially- and nano angle-resolved photo emission spectroscopy (nARPES) help to give a broad range of information both on the electronic band structure of the system and the nature of the defects\cite{10.1021/acsnano.8b06574}. Non-contact atomic force microscopy (ncAFM) further enables structure identification of 1DMs that host a TLL, where a metallic tip is functionalized with a CO molecule\cite{Barja2016,10.1021/acsnano.9b04611}. 

We report the role of the graphene substrate in TLL formation within TMD 1DM heterostructures. Two main types of 1DM defects are identified, in addition to combinations and strained subsets, where their controlled introduction enables TLL formation. Exclusive doping (charge transfer) from graphene into a 1DM gives on the order of $\sim$3-8 electrons per measured 1DM, which brings the E$_F$ of graphene near the Dirac point, reducing its screening power and, therefore, increasing the electron-electron interaction in the 1DMs. We also observe an atomically localized band gap renormalization over the 1DM/TLL systems, with indications of the conduction band taking part in TLL formation. Additionally, as further proof that graphene is in fact playing a fundamental role and bolstering our hypothesis, when 1DMs are in contact with a wide band gap material, such as multilayered TMD atop graphene, the TLL structure does not appear. These findings underscore the unique role of graphene as an ideal substrate for TLL formation, striking a balance between charge donation and minimal screening, where, in comparison, metals over-screen electron-electron interactions and gapped systems lack sufficient free carriers. This insight opens new avenues for engineering 1D correlated states in 2D materials.

\section*{Results}

\subsection*{Defect Creation}

In order to study 1DM defects that may host a TLL, we perform STM/STS studies on both unmodified WS$_2$, grown epitaxially via chemical vapor deposition (CVD) on a graphene/SiC substrate, and the same sample after Ar$^+$ sputtering and annealing in-situ to induce defectivity (Fig. \ref{fig1:topo} (a))\cite{Kastl_2017,Kastl_2018,C7NR02025B}. Comparative results of low-energy sputtering in contrast to a low-temperature anneal are shown (Fig. \ref{fig1:topo} (b-c)). As-grown samples are not significantly modified by annealing up to 250 $^\circ$C, while chalcogen vacancies (V$_{\rm S}$) start to form at 600 $^\circ$C\cite{PhysRevLett.123.076801}. Low-energy Ar$^+$ sputtering at chalcogen creation sample temperatures ($SA_{step}$) greatly increases the density of V$_{\rm S}$ and 1DM defects (see Supplementary Notes 1 and 2 and Supplementary Figs. 1 and 2 for both density calculations and SRIM simulations) compared to only annealing. Post $SA_{step}$ annealing at 600 $^\circ$C ($SAA_{step}$) substantially reduces point defect density from the $SA_{step}$ [0.116 to 0.220 V$_{\rm S}$/nm$^2$] to the $SAA_{step}$ [0.001 to 0.007 V$_{\rm S}$/nm$^2$], where 1DM defects with increased length [L$_{SA_{step}}$ = 3.37 $\pm$ 2.87 nm, L$_{SAA_{step}}$ = 8.75 $\pm$ 4.01 nm] are formed (see Supplementary Fig. 1). Densities of 1DM defects are also reduced, to a lesser extent compared to V$_{\rm S}$, from the $SA_{step}$ [0.012 to 0.030 defect/nm$^2$] to the $SAA_{step}$ [0.008 to 0.010 defect/nm$^2$], as 1DM defect length is increased via annealing. If multiple $SAA_{step}$ cycles are performed, longer 1DM defects [L$_{SAA_{3 x step}}$ = 12.95 $\pm$ 5.01 nm] can form (see Supplementary Fig. 3). After sputtering the sample for longer than 2 minutes, substantial degradation that is accompanied by an electronic structure change becomes measurable (see Supplementary Fig. 3 (e)).

We make use of ncAFM with a functionalized CO tip to identify chalcogen vacancy lines, strained 1DMs resembling 4|4P or 4|4E intermediate defects, and combinations of chalcogen vacancies with strained 1DM defects (see Supplementary Figs. 4 and 5). These defects are characterized by their distinct electronic behaviors—most exhibit electron-like characteristics, while other 1DMs demonstrate hole-like behavior (see Supplementary Fig. 6). This classification is based on the modulation of electron density with bias\cite{PhysRevX.9.011055,Zhu2022}. Notably, the band dispersion observed in these defects mirrors that of known MTB structures\cite{Batzill_2018,10.1021/acsnano.5b00410,10.1002/aelm.201600468}. For the hole-like behavior, our findings align with the 4|4E MTB similar to what was reported by Jolie et al., and for the electron-like modulation, we observe this behavior in chalcogen vacancy lines and strained 1DMs resembling the 4|4P MTB found by Zhu et al. All structures display features of a TLL\cite{PhysRevX.9.011055,10.1021/acsnano.0c05397,Zhu2022}. In addition to defect structures measured with ncAFM, we also show a power-law dependence in as-acquired STM images in single-line defects and fully formed 60$^\circ$ MTB formations, where this behavior (characteristic of a TLL) is not present in structures that instead contact an underlying layer of WS$_2$ (see Supplementary Fig. 7). An isolated 1D MTB in an otherwise unaltered MX$_2$ lattice would be highly strained. If isolated structures are formed, additional local strain-relieving features, such as Mo chains or chalcogen vacancies, can appear at either their endpoints or along their length, as demonstrated in Supplementary Fig. 5. This observation highlights a defect at the 1DM endpoint, where we measured a strained 4|4P intermediate symmetry thereafter. Our analysis reveals that among the types of 1DM defects measured, the isolated strained 1DM structures resemble intermediate defects in the process of forming fully relaxed MTB structures from V$_{\rm S}$ defects\cite{Wang2022,10.1021/acsnano.5b00554,10.1021/acsnano.8b01610}. Chalcogen vacancy lines, however, create less lattice strain of the isolated 1D defects measured with a single $SAA_{step}$ cycle, and likely represent the majority of defects created. While 1DMs have been more extensively studied in MoS$_2$ and MoSe$_2$, high formation energy in WS$_2$ has hindered atomic-scale investigations. We are able to take advantage of sulfur reduction techniques coupled with tandem atomic-scale measurements. Upon further investigation with STS, the two types of defects introduced during both $SA_{step}$ and $SAA_{step}$ are verified to be 1DMs and V$_{\rm S}$, as detailed in Fig. \ref{fig1:topo} (d). Here, the band gap of surrounding WS$_2$ is measured to be on the order of 2.5 eV and V$_{\rm S}$ shows deep unoccupied defect states split by spin-orbit coupled W \emph{d} states, which is consistent with previously measured values\cite{PhysRevLett.123.076801}. Conversely, 1DMs show an energy gap (E$_{\rm gap}$) between the HOS and LUS around the E$_F$ that is much smaller compared to that of WS$_2$, and is dependent on 1DM length. An important finding is 1DMs engineered into bilayer and multilayer WS$_2$ (grown on the same graphene/SiC substrate) do not exhibit any sign of a TLL (shown in the spectra of Fig. \ref{fig1:topo} (d) and Supplementary Fig. 7). This suggests a fundamental role of the substrate in the formation of a TLL. The HOS-LUS position around the E$_F$ can shift and is driven by, e.g., neighboring defects, tip-induced effects, and substrate variation\cite{PhysRevLett.123.076801,Zhu2022}. Moreover, analogous to the discerned power-law behavior, no small gap is observed in defects formed on multilayer WS$_2$ (Fig. \ref{fig1:topo} (d)). Evidently, one key ingredient for the formation of a TLL in 2D material heterostructures is the nearest vertical contact. This motivated the detailed study of the local and macroscopic electronic structure to unveil the mechanism behind TLL formation in WS$_2$ 1DMs that are in contact with graphene.

\subsection*{Spatially Dependent Electronic Structure}

In order to better investigate the physics of the 1DM bandgap formed in WS$_2$, we make use of both point STS and differential conductance mapping, which are powerful tools for screening defects\cite{Barja2016,PhysRevLett.123.076801,Thomas2022}. Fig. \ref{fig2:spec} (a) showcases dense STS collected along an electron-like 1DM to extract behavior above and below the measured E$_{\rm gap}$, where representative dispersion linescans of the HOS ($\psi_{-}$) and the LUS ($\psi_{+}$) are depicted in Fig. \ref{fig2:spec} (b-c). The number of quantum-well nodes changes as a function of bias, where the number of nodes increases in an integer fashion as the bias is ramped above and below the HOS and the LUS at -0.032 V and 0.025 V in Fig. \ref{fig2:spec} (b), respectively. This 1D particle-in-a-box behavior, which has been demonstrated for both electron-like and hole-like 1DMs in MoSe$_2$ and MoS$_2$, shows an increasing number of nodes moving further past the LUS (with decreasing period) and a decreasing number of nodes moving below the HOS (with increased period) for the electron-like 1DM or the reverse behavior for the hole-like 1DM\cite{PhysRevX.9.011055,Zhu2022}. This is consistent with our measurements shown in Supplementary Fig. 6 and contrasts electron-like behavior (the predominant 1DM defect measured) to hole-like behavior. We find that the nodal periodicity at the HOS and LUS for a measured in-phase electron-like 1DM structure (Fig. \ref{fig2:spec} (a)) is 2.4 nm, and, for a 1DM showing out of phase behavior at the HOS and LUS (Supplementary Fig. 8), the periodicity at the HOS is near 3 nm and 2.4 nm at the LUS. These values are far above the lattice constant of 0.315 nm. We further outline the number of nodes as a function of length in Supplementary Fig. 8, where a single node is measured per 2.7 $\pm$ 0.6 nm defect length at the HOS. The measured periodicity for the inspected hole-like 1DM structure is 0.5 nm (Supplementary Fig. 6 (e-h)) at both the HOS and the LUS, which is not within an integer relationship with the lattice parameter of WS$_2$. Supplementary Figs. 9 and 10 further show conductance maps as a function of energy below the HOS and above the LUS for both structures, where the presence of a neighboring V$_{\rm S}$ scatters quantum-well states above the LUS and the number of observable nodes from the HOS to the LUS increases from 4 to 6 nodes, respectively, and this then increases from 6 at the LUS up to 10 nodes at 0.4 eV before nearing the conduction band minimum (CBM) of WS$_2$ in Supplementary Fig. 9. The dense (1$\times$128$\times$400) linescan of point spectroscopy captured in Fig. \ref{fig2:spec} (b) showcases expected spin and charge separation above and below the E$_{\rm gap}$. As low-energy excitations are not Fermi liquid quasiparticles in TLL theory, the spin- and charge-density waves exhibit different dispersions and velocities,  denoted as $\upsilon_{s}$ and $\upsilon_{c}$, respectively. Their ratio can be experimentally measured in the FT-STS measurement as \begin{math}K_{\rm c} = \upsilon_{s} / \upsilon_{c}\end{math}, where K$\rm _{c}$ is the Luttinger parameter. From our FT-STS shown in Fig. \ref{fig2:spec} (d), we extract a K$\rm _{c}$ of 0.47, which is near previously acquired values on other TMD systems\cite{PhysRevX.9.011055,Zhu2022}. E$_{\rm gap}$ as a function of length is shown in Supplementary Fig. 11, where E$_{\rm gap}$ is dependent upon length ($L$), scaling linearly with $L^{-1}$. This behavior stands in contrast to Peierls instability, where the E$_{\rm gap}$ is constant in the CDW case and does not exhibit a length dependence\cite{PhysRevX.9.011055}. We also look at the DOS measured in topographic STM images, as created defects are, on average, below the size regime shown to yield power-law scaling in point LDOS spectra\cite{Meden2000,10.1021/acsnano.0c05397}. TLL nodal oscillations are indistinguishable above the CBM of WS$_2$, however, a measure of signal intensity as a function of distance is feasible. Here, electron density is expected to decay according to K$\rm _{c}$ as \begin{math}\rho \sim x^{-K_{c}} \end{math}\cite{10.1021/acsnano.0c05397}, where the DOS ($\rho$) measured under constant-current as a function of distance (x), in nanometers, is shown in Supplementary Fig. 7. Here, constant-current line profiles are extracted along multiple electron-like 1DM defects (sample bias = 1.2 V) which yield a power-law exponent value of 0.49 $\pm$ 0.16 that is near the value obtained in the FT-STS and in agreement with previously measured 1DMs that host a TLL\cite{PhysRevX.9.011055,10.1021/acsnano.0c05397,Zhu2022}. We identify the TLL properties of these defects by observing a Luttinger parameter near 0.5, measuring an E$_{\rm gap}$ opening near E$_F$, identifying an E$_{\rm gap}$ dependence on 1DM length, visualizing 1D particle in a box behavior, and extracting evidence of spin-charge separation (See Supplementary Note 3). In previous studies, the observation of identical spectroscopic characteristics led to the determination of TLL within MTBs formed in other TMDs\cite{PhysRevX.9.011055,10.1021/acsnano.0c05397,Zhu2022}. We summarize the spectroscopic characteristics of chalcogen vacancy lines, strained 1DM structures, and a fully formed MTB (formed across WS$_2$ edges) in Fig. \ref{fig3:structure_compare}. All of these defects exhibit metallic characteristics within WS$_2$ and the capability for hosting a TLL, which contributes to the local modification of the electronic density of states. The significance of the underlying contact in the electronic structure formation is discussed in the next section.

\subsection*{Correlating Angle-Resolved Spectroscopy and Local Density of States}

We next perform nARPES to directly visualize the crystal band dispersion. The sub-$\mu$m probe in nARPES offers a spatial resolution capable of capturing the local inhomogeneity in the sample. 
Fig. \ref{fig4:arpes} showcases the as-measured band structure of the unmodified crystal versus the sample exposed to Ar$^+$ bombardment, where the sample was transferred from the STM chamber, after $SA_{step}$, in a nominally inert environment to the nARPES chamber with an additional annealing step before nARPES acquisition. The two spectra are collected from the same sample, where a small region is found to be unaffected by the $SA_{step}$, due to sample holder shadowing during preparation. The spectrum obtained from the non-defective structure is displayed in Fig. \ref{fig4:arpes} (a). It is collected along the $\Gamma$-K direction. Graphene and WS$_2$ keep epitaxial registry, therefore WS$_2$ also has the same crystal orientation\cite{C7NR05495E}. A sketch of the two Brillouin zones (BZ) (graphene in black, WS$_2$ in green) is highlighted in the inset of Fig. \ref{fig4:arpes} (a). Graphene exhibits sharp bands with the E$_F$ $\sim$400 meV above the Dirac point. This is consistent with graphene prepared from thermal decomposition of SiC, where the carbon-rich buffer layer between graphene and SiC substrate creates an electric dipole at the interface affecting the chemical potential of graphene\cite{PhysRevLett.108.246104}. Such a native gating also affects the WS$_2$ band, whose top of the valence band maximum (VBM) appears to be $\sim$1.48 eV below the E$_F$\cite{C7NR05495E}. The local maximum at $\Gamma$ appears below the maximum at K, confirming the ML nature of the TMD\cite{PhysRevB.83.245213}. The introduction of 1DMs deeply affects the bands of WS$_2$. Fig. \ref{fig4:arpes} (b) shows the spectrum obtained from the region with high defect density. A substantial band gap renormalization is observed, where both the magnitude and chemical potential position are affected. The gap renormalization arises as the self-energies of the band-edge states shift due to the Coulomb interaction among free carriers\cite{PhysRevB.47.9615}. The presence of 1DMs screens the Coulomb interaction in the WS$_2$ crystal leading to a smaller TMD gap. The HOS/LUS states, observed in STS, are not visible via nARPES experiment. The disorder dictated by the presence of defects results in a higher background signal and broader WS$_2$ bands. In order to have a clear signal of the HOS-LUS states, very ordered 1DMs with homogeneous length and orientation are necessary\cite{Ma2017}. The WS$_2$ occupied electron bands are shifted upwards by $\sim$500 meV. The position of the VBM is further confirmed by the spectrum collected with linear vertical polarization (see Supplementary Fig. 12). The graphene bands are also shifted up, although not by the same magnitude, with a subsequent change in the doping level. This is clear analyzing the Dirac bands collected near the K point of the graphene BZ along the direction highlighted in red within the inset of Fig. \ref{fig5:overview} (a,b). The momentum distribution curves (MDCs), collected at the E$_F$ (Fig. \ref{fig5:overview} (c,d)) can be fit with two Lorentzian functions. The position of their peaks defines a distance that approximates the diameter of the circle fitting the Fermi surface of graphene. Using Luttinger theorem\cite{Bostwick2007,PhysRevLett.125.176403,10.1021/acs.nanolett.3c01974}, it is possible to extract the doping level being n$_p = 1.2\times 10^{13}$ $\textrm{cm}^{-2}$ and n$_d=2.5\times 10^{12}$ $\textrm{cm}^{-2}$ for the as-grown and defective crystal, respectively (Fig. \ref{fig5:overview}). The unmodified doping level agrees well with the value found in literature\cite{Bostwick2007}. From these values and locally-measured defect densities (Supplementary Fig. 1 and Supplementary Note 2), we are able to determine that each 1DM hosts 3-8 electrons (0.021 $\pm$ 0.009 \begin{math}\frac{1DM}{nm^2}\end{math}). Our Monte Carlo calculations further suggest that our sample treatment does not have a direct impact on graphene (Supplementary Fig. 2). As a matter of fact, the presence of defects in graphene would open a gap at the Dirac point caused by an alteration of the system symmetry rather than shifting its chemical potential\cite{PhysRevB.101.235116}. The shift of the Dirac point in graphene is the result of charge transfer from graphene to the newly formed 1DMs in WS$_2$. We can exclude a charge transfer involving the single V$_{\rm S}$ having its characteristic in-gap state above Fermi level and therefore unoccupied, as displayed in the STS curve in Fig. \ref{fig1:topo} (d). Our findings suggest that graphene plays a significant role in the formation of a TLL, both donating charge to the newly formed defects and providing a weaker electronic screening due to the lower carrier density near the neutrality point. This, overall, increases the electron-electron interaction strength in the TMD. 

An analysis of the energy distribution curves (EDC) intersecting the VBM is presented in Fig. \ref{fig4:arpes} (d), denoted by the vertical dashed line in Fig. \ref{fig4:arpes} (a-b). The EDCs are compared with STS curves obtained from the unmodified crystal and a 1DM formed on TMD. The pristine structure (shown as blue lines) exhibits a satisfactory alignment at the expected onset of the VBM. The horizontal semitransparent bars, displayed in blue and red, highlight the energy region associated with the VBM onset for both the pristine and defective systems. A distinct kink in the EDC indicates the onset of the VBM as extracted from the nARPES data, marked by the black arrow. The observed band shift is primarily attributed to the band gap renormalization caused by the presence of the 1DM, in contrast to isolated V$_{\rm S}$. Specifically, the onset of the VBM indicated by the STS curve obtained from the V$_{\rm S}$ (represented in Fig. \ref{fig1:topo} (d)) does not coincide with the onset of the VBM extracted from the EDC in Fig. \ref{fig4:arpes} (b). However, the STS curve obtained from the 1DM (illustrated by the dark red line) exhibits the onset of the VBM that aligns with the EDC derived from the nARPES data acquired after the annealing step (depicted by the light red line). This behavior is additionally measured with spatially-localized STS acquisition in Fig. \ref{fig4:arpes}, where a shift in the VBM is seen approaching a 1DM with an STM tip. The measured onsets of the VBM (-1.16 $\pm$ 0.04 eV) and the CBM (0.74 $\pm$ 0.09 eV) were measured locally with STS, which corresponds with the appearance of the HOS and LUS. The relative shift of the VBM is +530 meV (CBM is shifted by -80 meV) and is depicted in Fig. \ref{fig4:arpes} (c), which matches well to the band structure acquired by nARPES. Spatially-resolved STS spectra reveal a band bending across the 1DM, akin to results by Murray et al.\cite{10.1021/acsnano.0c04945} and the pristine WS$_2$ that extends over 500 pm, giving rise to a sharp junction across the two regions. In addition, W 4$f$ and S 2$p$ core levels in Supplementary Fig. 13 show both core levels are rigidly shifted to lower binding energy in a similar fashion as the bands in the valence region. This suggests that the band gap renormalization is overall electrostatically driven. W and S peak fitting upon 1DM formation displays a weaker component that is further shifted towards a lower binding energy of an additional few hundreds of meV due to the local chemical environment. Both nARPES measurements and STS highlight the importance in choice of substrate to determine the electronic properties of 1DM defects in TMDs. A screening effect is caused by graphene, which occurs due to the depletion of electrons in favor of 1DM states that enhances electron-electron interaction within the defect. This interaction leads to a metallic behavior that ultimately influences the renormalization of the band-gap in WS$_2$.

As anticipated above, graphene itself plays an important role in the formation of a TLL. In order to better understand its influence, we also examine the behavior of 1DMs in multilayered WS$_2$ on a graphene/SiC substrate (shown in Fig. \ref{fig1:topo} (d) and Supplementary Figs. 7, 14, and 15) to compare direct WS$_2$/graphene contact against defects with greater separation to graphene. Defects shown in Supplementary Fig. 14 (a) are over 3 ML of WS$_2$, which is confirmed by STS in Supplementary Fig. 14 (b) and in agreement with similar measured WS$_2$ systems\cite{Cochrane2021}. Dense STS is acquired over a 1DM in Supplementary Fig. 15, where there are consistently no in-gap states, or evidence of a TLL, across the entirety of the defect. Additionally, there are quantum-well states that form above the CBM of WS$_2$, which is in stark contrast with a 1DM in direct contact with graphene. As the electrons from the tip are tunneling into unoccupied 1DM states at positive sample bias with no reduced screening effects from the underlying substrate, the result is a 1D Fermi liquid. This behavior is further verified in constant-current line profiles (Supplementary Fig. 7) over multiple defects within a multilayered WS$_2$ system. Contrary to 1DMs with direct contact to graphene, power-law fittings produce an exponential parameter of 0.99 $\pm$ 0.15 at a sample bias of 1.2 V. This has been shown previously with a 1DM over Au\cite{10.1021/acsnano.0c05397}, where line profiles also decayed as a Fermi liquid. The presence of a 1D Fermi liquid in 1DMs with reduced proximity versus the presence of a TLL in 1DMs with direct graphene contact further highlights the importance of the underlying substrate. Cross-correlated measurements made in this report are able to directly visualize the E$_F$ position related to the presence of 1DMs over graphene.

\section*{Discussion}

This work presents a new and controllable way to create 1DMs in WS$_2$ epitaxially grown on graphene. We show how 1DMs host a TLL, in which spectroscopic and topological signature is shown via STM/STS and ncAFM. A band gap opening is observed near the E$_F$ confirming the correlated nature of the electronic state inside the 1D defect. We demonstrate the formation of electron-like and hole-like 1DMs with combined STM/STS in WS$_2$ that display similar spectroscopic features but a distinct spatial and energetic difference in conductance image mapping. This behavior is due to respective electronic characteristics, which confirms earlier reports\cite{10.1021/acsnano.0c05397,PhysRevX.9.011055,Zhu2022}. Data obtained with ncAFM paired with STM/STS also confirms the formation of a TLL within WS$_2$ fully formed MTBs, intermediate 1DM formations, and chalcogen vacancy lines. By means of nARPES, we were able to correlate scanning probe spectroscopic 1DM features to the band structure of the crystal. We observed how defective states behave as an acceptor of graphene electrons and that they cause a massive band gap renormalization of WS$_2$, where the presence of 1DMs screen the Coulomb interaction of WS$_2$ carriers leading to an overall smaller TMD gap. The same effect is reflected on core levels where we observe a similar chemical shift in binding energies. We also compare and contrast 1DM behavior directly over graphene to that of a 1DM with increased separation that is instead contacted to another layer of WS$_2$. Here, the effective electronic structure measured between STS and nARPES demonstrates that the direct heterostructure of graphene with a 1DM embedded into WS$_2$, and the subsequent induced charge transfer into the 1DM, is the driving mechanism behind the formation of TLLs and is critical in these types of systems.

The unique role of graphene is further highlighted by its ability to provide charge carriers while maintaining weak screening, a balance that is absent in either metallic or fully gapped substrates. Unlike metals, which over-screen electron-electron interactions, or semiconductors, which do not contribute sufficient free carriers, graphene's semi-metallic nature enables both charge transfer and the necessary electron-electron interactions for TLL formation. This demonstrates the fundamental importance of choosing the right substrate in engineering strongly correlated 1D electronic states.

Quasi-particle formation within 1D structures can have immediate relevance in quantum information processing, where application of such materials has yet to be manifested in functional devices. Additionally, electron transport across semiconductor to metal transitions may be beneficial in ultrafast electronic systems. Here, we provide a step-by-step approach to produce 1D metallic structures within a 2D semiconducting material, which holds relevance in atomic-scale tailorable systems and electronic modification at the nanoscale. We further anticipate engineered TMD materials to be relevant in spin-polarized measurements, charge state effects, and spin transport. 

\section*{Methods}
\subsection*{Scanning Probe Microscopy Measurements}
All measurements were performed with a Createc GmbH scanning probe microscope operating under ultrahigh vacuum (pressure < 2x$10^{-10}$ mbar) at liquid helium temperatures (T < 6 K). Either etched tungsten or focused ion beam cut platinum iridium tips were used during acquisition. Tip apexes were further shaped by indentations into a gold substrate. STM images are taken in constant-current mode with a bias applied to the sample. STS measurements were recorded using a lock-in amplifier with a resonance frequency of 683 Hz and a modulation amplitude of 5 mV. FT-STS fit lines were derived from local extrema on spin and charge branches in accordance to analysis performed in Zhu et al\cite{Zhu2022}. A padding of 10 pixels was used to showcase the region of separation.

In ncAFM measurements, a qPlus quartz-crystal cantilever was used (resonance frequency, f$_0$ $\approx$ 30 kHz; spring constant, k $\approx$ 1800 N/m; quality factor, Q > 18,000; and oscillation amplitude, A $\approx$ 1 {\AA})\cite{Schuler2020-fh}. The metallic tip was functionalized with a CO molecule for enhanced resolution\cite{10.1021/acsnano.9b04611}.

\subsection*{Angle-Resolved Photoemission Spectroscopy Measurements}
ARPES experiments were performed in ultra high vacuum at T= 6K at beamline 7.0.2 (MAESTRO) at the Advanced Light Source. The beam-spot size was $\approx$ 1 $\mu$m. The photon energy for the valence band structure was $h\nu$ = 150 eV and $h\nu$ = 350 eV for core levels. The XPS curve-fitting analysis was performed using a convolution of Doniach-Sunjic and Gaussian line shapes superimposed on a background built of a constant, a linear component, and a step-function. For each S 2$p$ spin-orbit doublet, a spin-orbit splitting of 1.2 eV and a branching ratio I(2$p$3/2) : I(2$p$1/2) = 2 : 1 (defined in terms of peak areas) were used. The W 4$f$ spin-orbit doublets were fit using individual components with a spin-orbit splitting of 2.2 eV, to take into account non-linearities.

\subsection*{Sample Preparation}
Islands of WS$_2$ were grown on graphene/SiC substrates with an ambient pressure CVD approach. An MLG/SiC(0001) substrate with 10 mg of WO$_3$ powder on top was placed at the center of a quartz tube, and 400 mg of sulfur powder was placed upstream. The furnace was heated to 900 $^{\circ}$C and the sulfur powder was heated to 250 $^{\circ}$C using a heating belt during synthesis. A carrier gas for process throughput was used (Ar gas at 100 sccm) and the growth time was 60 min. The CVD grown WS$_2$/MLG/SiC was further annealed in vacuo at 400 $^{\circ}$C for 2 hours. 

WS$_2$ was sputtered with an argon ion gun (SPECS, IQE 11/35) that operated at 0.1 keV energy with 60$^{\circ}$ off-normal incidence at a pressure of 5$\times$10$^{-6}$ mbar and held at 600 $^{\circ}$C. A rough measure of current (0.6$\times$10$^{-6}$ A) enabled the argon ion flux to be estimated at (1.5$\times$10$^{13}$ \begin{math}\frac{ions}{cm^2s}\end{math}), where sample irradiation cycles spanned up to 30 seconds.

Samples were transferred from the STM to the nARPES chamber using an Ar (low-oxygen) suitcase to enable cross-correlative studies with minimal sample degradation risk. Samples were then annealed for 12 h at 250 $^{\circ}$C and transferred to a 6 K sample stage for nARPES data acquisition.
\section*{Data Availability} All data needed to evaluate the conclusions exhibited are present in the paper and/or the supplementary information. The data generated for spatially-resolved STS (Fig. \ref{fig2:spec} (b)) in this study are provided in the Source Data file. Other data are available upon request.
\section*{Code Availability} Software used for analysis are either presented in the supplementary information or can be provided upon request.

\section*{Acknowledgements}
The authors thank Alexander Stibor and John Turner, from the Molecular Foundry and the National Center of Electron Microscopy, and also Nino Hatter and Sebastian Baum, from CreaTec Fischer \& Co. GmbH, for helpful discussions and experimental support. This material is based upon work supported by the U.S. Department of Energy, Office of Science, National
Quantum Information Science Research Centers, Quantum Systems Accelerator. Additional support is
acknowledged from the Center for Novel Pathways to Quantum Coherence in Materials, an Energy Frontier Research Center funded by the U.S. Department of Energy, Office of Science, Basic Energy Sciences. Work was performed at the Molecular Foundry and at the Advanced Light Source, which was supported by the Office of Science, Office of Basic Energy Sciences, of the U.S. Department of Energy under contract no. DE-AC02-05CH11231. J.C.T., A.R., and A.W.-B acknowledge support from the U.S. Department of Energy, Office of Science, Basic Energy Sciences in Quantum Information Science under Award Number DE-SC0022289. S.K and J.A.R. acknowledge support from the National Science Foundation Division of Materials Research (NSF-DMR) under awards 2002651 and 2011839. J.T.K and F.A. acknowledge financial support by the Deutsche Forschungsgemeinschaft (DFG) through the TUM International Graduate School of Science and Engineering (IGSSE), GSC 81.
\section*{Author Contributions Statement}
A.R., J.C.T., and A.W.-B. conceived and carried out the experiments. A.R., J.C.T., and J.T.K carried out nARPES/XPS measurements with the assistance of C.J., A.B., and E.R. E.B. and J.C.T. contributed to SRIM/TRIM simulations. A.R., J.C.T., and J.T.K. carried out ncAFM measurements with the assistance of H.-Z.T. and M.F.C. A.R., J.C.T., and J.T.K. performed all STM/STS experiments with additional contributions from A.R, E.W., A.S., D.F.O., F.A., W.A., and A.W.-B. A.R. and J.C.T. performed all nARPES related data analysis with support from C.J., A.B., and E.R. A.R and J.C.T. performed all ncAFM, STM, and STS related data analysis with support from A.R, E.W., A.S., D.F.O., J.B.N., M.F.C., F.A., W.A., and A.W.-B. Z.Y., T.Z., S.K., J.A.R. and M.T. synthesized the samples. All authors discussed the results and contributed towards the manuscript.  
\section*{Competing Interests Statment} The authors declare no competing interests. 


\renewcommand{\figurename}{Fig.}
\renewcommand{\thefigure}{1}
\begin{figure}[H]
    \centering
    \includegraphics[width = 1.0 \linewidth]{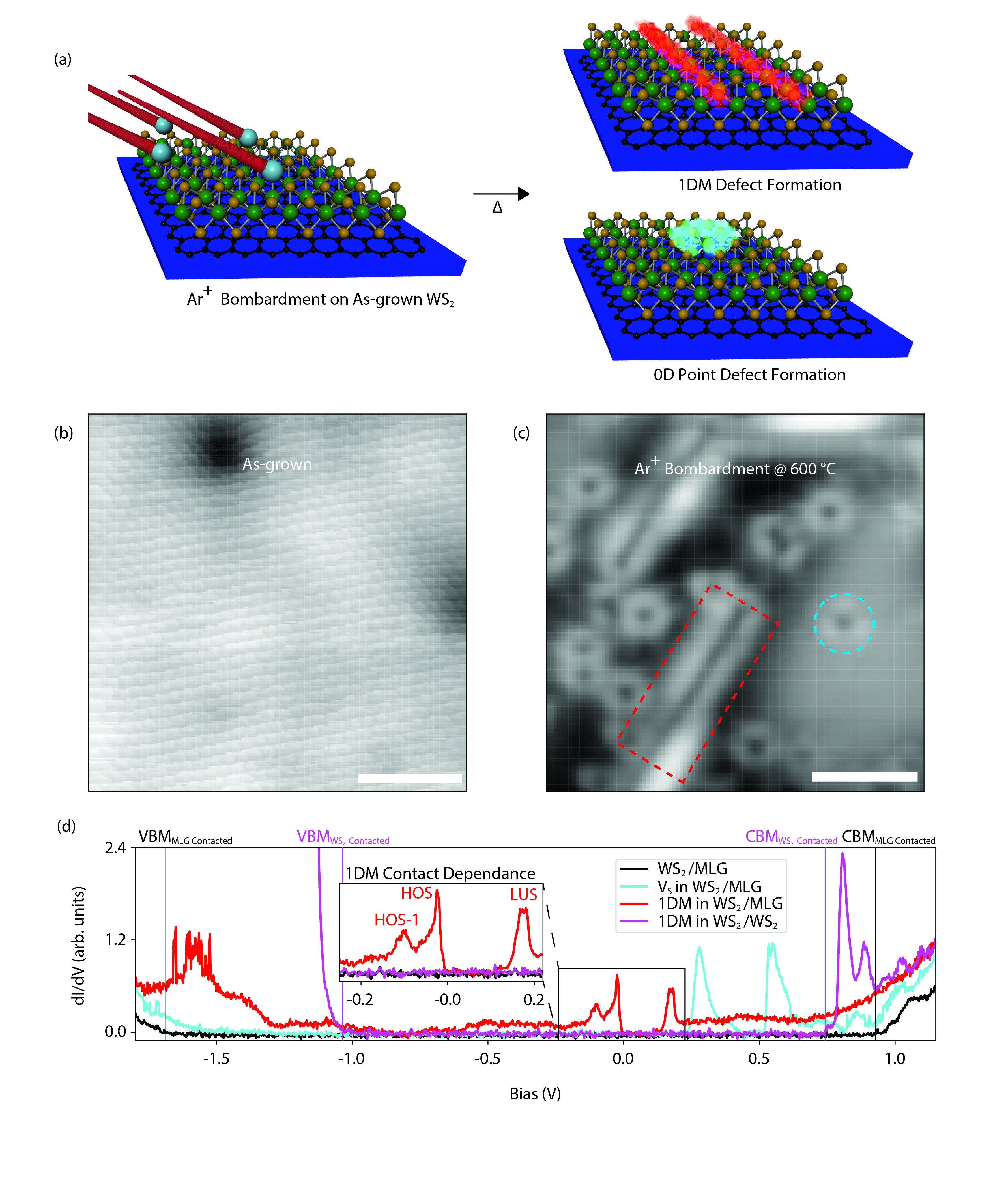}
    \caption{\textbf{WS$_2$ Defect Introduction.} (a) An overview of the defect creation process, where Ar$^+$ irradiation and annealing steps create both 0D and 1DM defects into an otherwise unmodified WS$_2$ monolayer. (b) Scanning tunneling micrograph depicting pristine WS$_2$ before Ar$^+$ bombardment ($I_{tunnel}$ = 30 pA, $V_{sample}$ = 1.2 V). (c) After the pristine sample is heated and exposed to an Ar$^+$ sputter ($I_{tunnel}$ = 30 pA, $V_{sample}$ = 1.2 V), both V$_{\rm S}$ and 1DMs are present. Scale bars, 2 nm. (d) Point spectroscopy, in the form of the LDOS, for unmodified WS$_2$ in contact with monolayer graphene (MLG), V$_{\rm S}$ within WS$_2$ and in contact with underlying MLG, a 1DM within WS$_2$ and in contact with MLG, and, for comparison, a 1DM within WS$_2$ instead contacted to underlying WS$_2$ (contact dependence is highlighted by the inset plot). States around E$_F$ for the MLG-contacted 1DM are labeled as HOS-1, HOS, and LUS. The valence band maximum (VBM) and conduction band minimum (CBM) are labelled with horizontal lines for unmodified WS$_2$ (black) and a 1DM in WS$_2$ contacted to underlying WS$_2$ (magenta).}
    \label{fig1:topo}
\end{figure}

\renewcommand{\thefigure}{2}
\begin{figure}[H]
    \centering
    \includegraphics[width = 1.0 \linewidth]{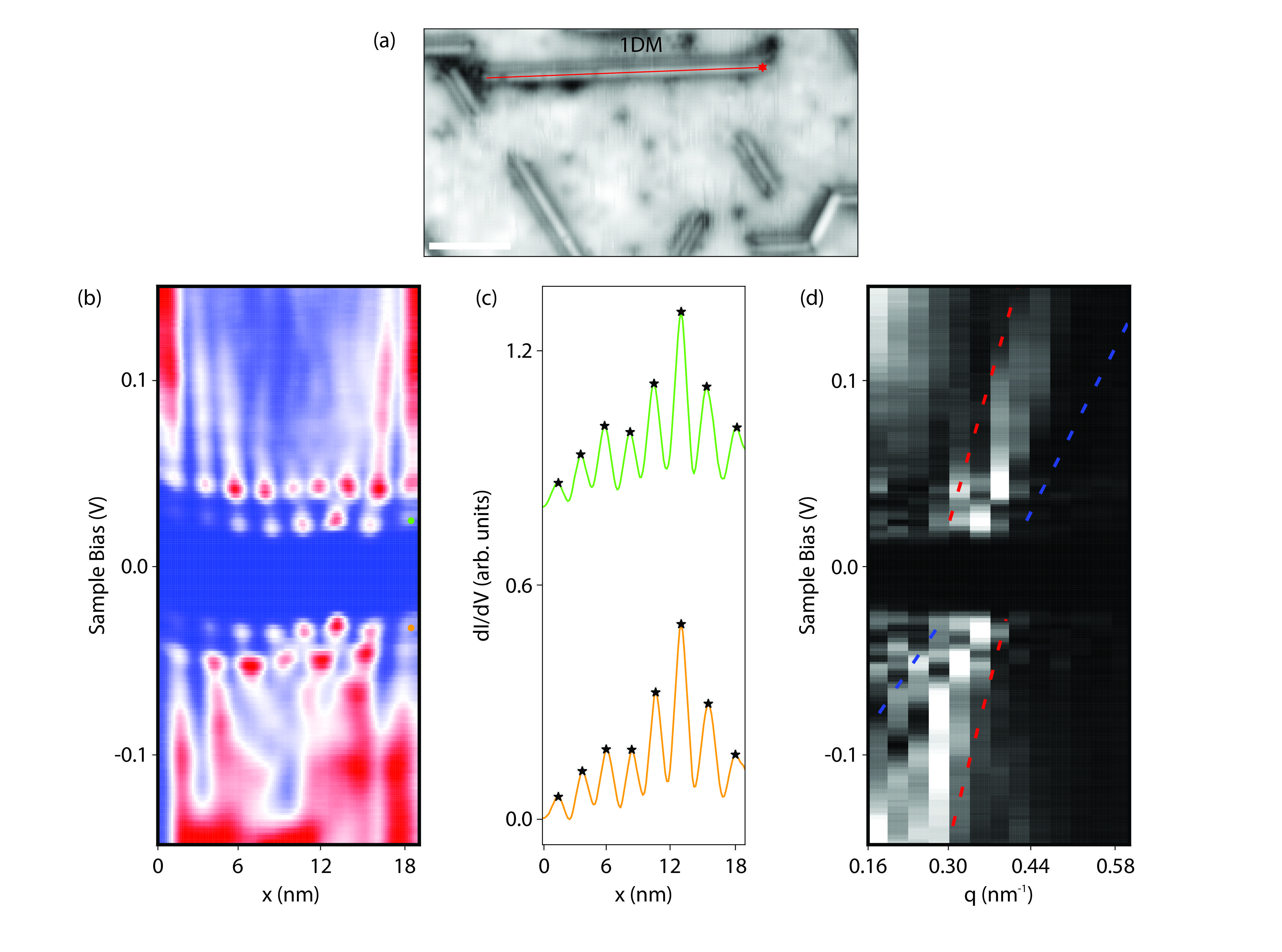}
    \caption{\textbf{Dense LDOS Mapping.} (a) An electron-like 1DM is measured topographically, where the red line shown depicts a length of 18.96 nm ($I_{tunnel}$ = 20 pA, $V_{sample}$ = 1.4 V). Dense LDOS spectra is collected along (a) an electron-like 1DM (1x128x400 pixels) ($V_{modulation}$ = 5 mV, $I_{set}$ = 150 pA) beginning at the starred point along the red line. This is shown as a (b) function of bias and distance. The HOS and LUS are identified (c) at -0.032 V (orange) and 0.025 V (green). A 1D particle-in-a-box behavior is present above and below the HOS and LUS (maxima labeled as starred points). The number of nodes (N) increases linearly (N+1) from the LUS to LUS+1 (8 to 9), and also decreases from the HOS to HOS-1 (8 to 7). An FFT of (b) is shown in (d), where a spin and charge separation onset is seen above and below the E$_F$, and both the spin (blue) and charge (red) branches can be monitored from -0.15 V to 0.15 V. A K$\rm _{c}$ value of 0.47 is extracted from (d).}
    \label{fig2:spec}
\end{figure}

\renewcommand{\thefigure}{3}
\begin{figure}[H]
    \centering
    \includegraphics[width = 1 \linewidth]{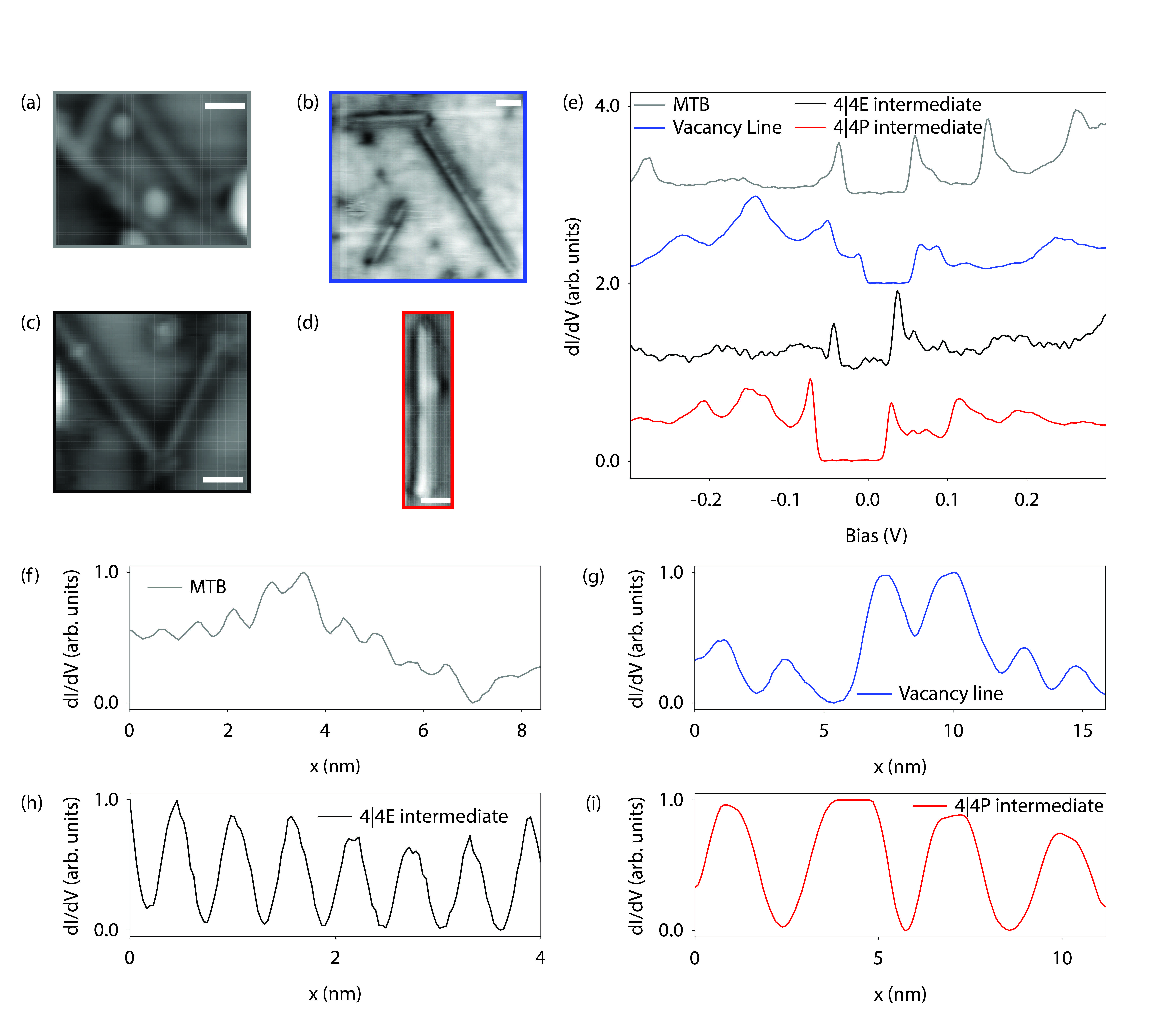}
    \caption{\textbf{Electronic Structure Comparison of Different 1DMs.} Constant current image over (a) a fully relaxed MTB across WS$_2$ edge regions ($I_{tunnel}$ = 30 pA, $V_{sample}$ = 1.2 V), (b) a vacancy line structure showing 120 degree rotation within WS$_2$ ($I_{tunnel}$ = 20 pA, $V_{sample}$ = 1.4 V), (c) a 4|4E intermediate structure ($I_{tunnel}$ = 30 pA, $V_{sample}$ = 1.2 V), and (d) a 4|4P intermediate 1DM ($I_{tunnel}$ = 30 pA, $V_{sample}$ = 1.2 V). Scale bars, 2 nm. (e) dI/dV spectra recorded over each 1DM defect case, where all show a measurable E$_{\rm gap}$ ($V_{modulation}$ = 5 mV) near the E$_F$. In addition, we highlight dI/dV linescans extracted from the HOS as a function of defect distance (f-i), which all exhibit similar oscillatory behavior.}
    \label{fig3:structure_compare}
\end{figure}

\renewcommand{\thefigure}{4}
\begin{figure}[H]
    \centering
    \includegraphics[width = 1.0 \linewidth]{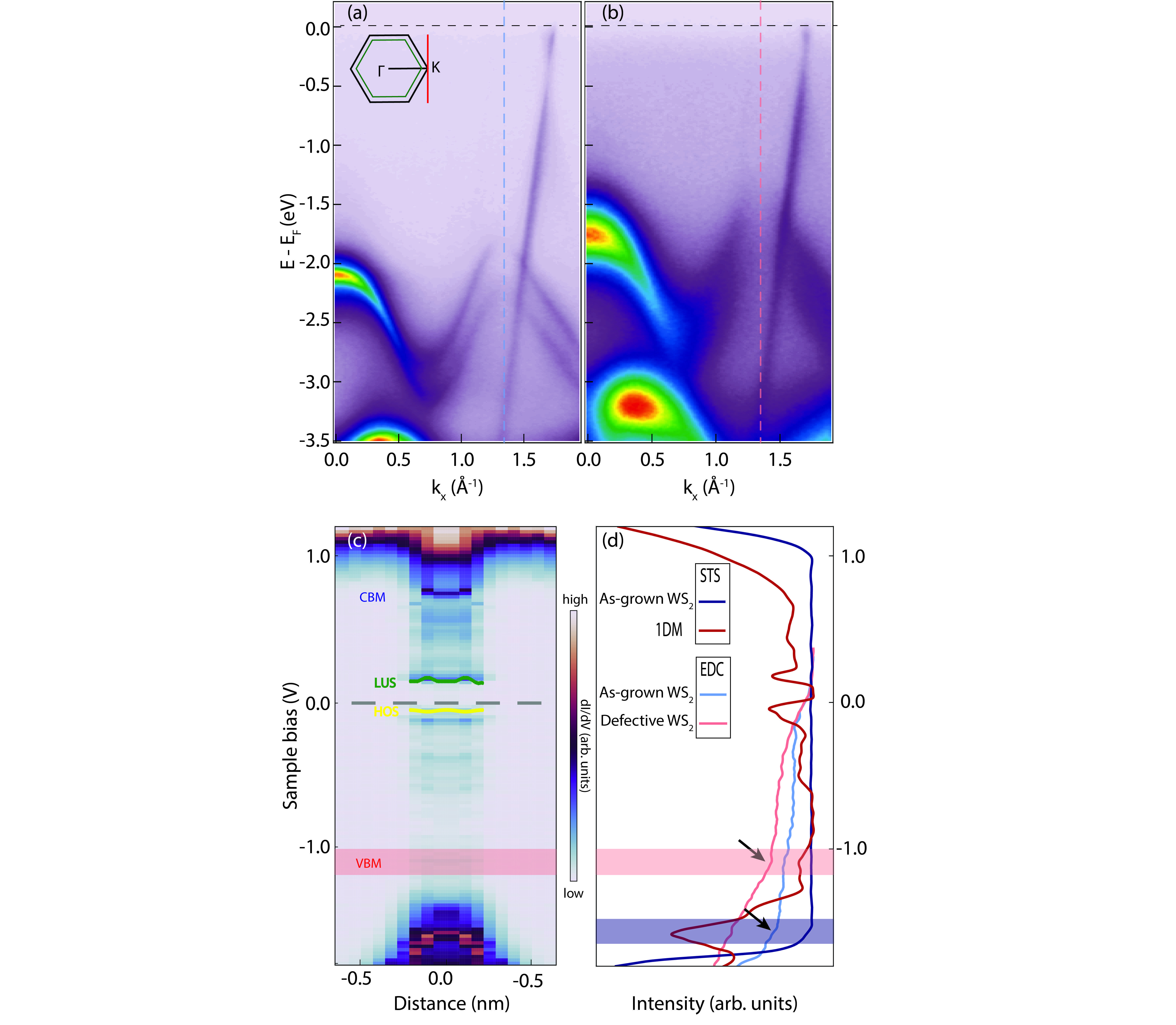}
    \caption{\textbf{nARPES Band Structure Comparison.} (a) Unmodified and (b) defective band structure of WS$_2$ on graphene. The inset displays the WS$_2$ (green) and the graphene (black) BZ. The spectra are collected along the $\Gamma$-K orientation. (c) LDOS spectra collected from pristine WS$_2$ to a 1DM ($V_{modulation}$ = 5 mV, $I_{set}$ = 150 pA) and then recorded in reverse. The as-measured VBM, CBM, HOS, and LUS are highlighted with relative positions to E$_F$. (d) EDCs overlapped with STS from unmodified and defective WS$_2$. Dark blue and dark red are the STS signals collected on pristine WS$_2$ and on a 1DM, respectively. The light blue and light red lines are the EDCs collected at the $K$ point of the BZ (dashed vertical lines in panels (a) and (b) respectively). Horizontal semitransparent bands display the onset of the VBM for pristine (blue) and defective (red), corresponding to EDC kinks clarified with black arrows.}
    \label{fig4:arpes}
\end{figure}

\renewcommand{\thefigure}{5}
\begin{figure}[H]
    \centering
    \includegraphics[width = 1.0 \linewidth]{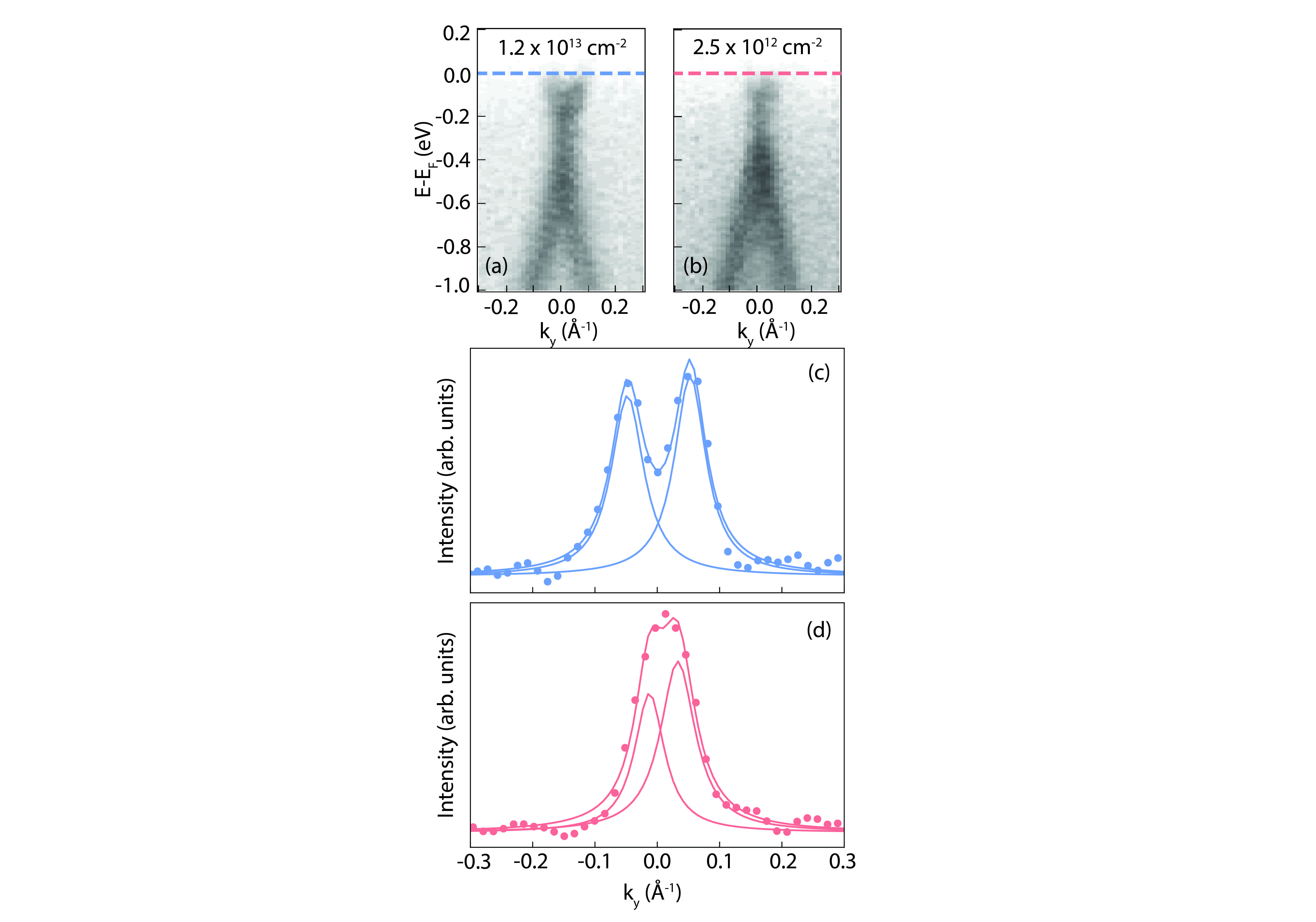}
    \caption{\textbf{Graphene Charge Transfer.} The Dirac cone shift from the Graphene bands is related to the number charges transferred from the substrate into 1DM defective regions, where defective regions are closer to the neutrality point. nARPES spectra displaying graphene bands for both (a) unmodified and (b) defective samples. The spectra are collected along the portion of BZ highlighted by the red line of the inset in Fig. 4 (a). Panel (c) and (d) display MDCs collected near E$_F$ for pristine and processed structure respectively. The distance between the two Lorentzian peaks fitting each curve approximates the diameter of the Fermi surface, which is proportional to the carrier density in graphene. A smaller diameter indicates lower carrier density, highlighting the charge transfer process when WS$_2$ 1DMs are present.}
    \label{fig5:overview}
\end{figure}

\end{document}


\maketitle
\setstcolor{red}
\section*{Supplementary Notes}
\subsection*{1| Ar$^+$ sputtering and SRIM simulations}
Monte Carlo simulations based on The Stopping and Range of Ions in Matter, SRIM simulations\cite{ZIEGLER20101818}, were used to evaluate preparation conditions using Ar$^+$ bombardment. The Transport of Ions in Matter (TRIM) calculation, which assumes amorphous targets, with 50,000 ions was determined to be sufficient for simulation convergence; between simulating 20,000 and 50,000 ions, the variation in the number of vacancies created was less than 2\% for all possible atomic vacancies. The Ar$^+$ energy was set to 0.1 keV to gauge angle dependence and fixed at an angle of 60$^{\circ}$ to gauge energy dependence, where these values were chosen to be used experimentally. We set the height of WS$_2$ to 0.72 nm, the height of graphene to 0.34 nm, and the height of SiC to 30 nm\cite{Mitterreiter2021-ih,Fox_2013}. Density, displacement energy, and surface binding energy values were matched to literature values\cite{PhysRevLett.109.035503,Susi2016,CHANG2014475} .

Using the estimated argon flux of 1.5$\times$10$^{13}$ \begin{math}\frac{ions}{cm^2s}\end{math} at 30 seconds of irradiation, we obtain a value of 4.5 \begin{math}\frac{ions}{nm^2}\end{math}. Each ion is predicted to induce 3 V$_{\rm S}$, or 13.5 \begin{math}\frac{V_{\rm S}}{nm^2}\end{math}. Considering the local measurement of 0.168 $\pm$ 0.052 \begin{math}\frac{V_{\rm S}}{nm^2}\end{math} and 1D metal (1DM) formation of 0.021 $\pm$ 0.009 \begin{math}\frac{1DM}{nm^2}\end{math} with a length of 3.37 $\pm$ 2.87 nm. We use simulation results to estimate defect creation and local measurements to approximate defect density. 
\subsection*{2| Defect analysis}
An object class can be instantiated within Python, where each image can be loaded for analysis with a given size (nm). Defects are selected initially by inspection, and local minima or maxima are calculated within a given pixel range. Each selected defect is then cross-checked and input into a graph, where density can be calculated by the number of defects within a the given area, assuming an equal N$\times$N image.
\begin{verbatim}
class node(object):
    def __init__(self,position,value):
        self.value=value
        self.position=position
    def getPosition(self):
        return self.position
    def getvalue(self):
        return self.value
    def getNodeHash(self):
        return hash(str(self.position)+str(self.value))
    def __str__(self):
        return str('Pos:'+str(self.position)+' Val:'+str(self.value))

class edge(object):
    def __init__(self,src,dest):
        self.src = src
        self.dest = dest
    def getSource(self):
        return self.src
    def getDestination(self):
        return self.dest
    def getWeight(self):
        return self.dest.getvalue()
    def __str__(self):
        return str(self.src.getPosition())+'-->'+str(self.dest.getPosition())

class emap(object):
    def __init__(self):
        self.edges = {}
    def addNode(self,node):
        if node in self.edges:
            raise ValueError('Duplicate node')
        else:
            self.edges[node]=[]
    def addEdge(self,edge):
        src = edge.getSource()
        dest = edge.getDestination()
        if not (src in self.edges and dest in self.edges):
            raise ValueError('Node not in graph')
        self.edges[src].append(dest)
    def getChildrenof(self,node):
        return self.edges[node]
    def hasNode(self,node):
        return node in self.edges
    def display(self):
        for i in self.edges:
            print(i)
    def getedgelen(self):
        return len(self.edges)
       
class defect_map(object):
    def __init__(self):
        self.len=0
        self.images = []
        self.imsize = []
        self.density = []  
    def addimage(self,im,siz):
        self.images.append(im)
        self.imsize.append(siz)
        self.len += 1
    def getlen(self):
        return self.len     
    def select_defects(self, win):
        for idx in range(0,len(self.images)):
            outpts = selectdefects(self.images[idx],win,idx)
            dList = emap()    
            nlist = []
            k = 0
            for x, y, z in outpts:
                mol=node([x,y],k)
                dList.addNode(mol)    
                nlist.append(mol)
                k += 1
            visited = []
            for i in nlist:
                visited.append(i)
                for j in nlist:
                    if j not in visited:
                        dList.addEdge(edge(i,j))
            self.density.append(dList.getedgelen())    
    def getdensity(self):
        tmp = []
        for i in range(0,len(self.density)):
            tmp.append(self.density[i]/(self.imsize[i]**2))
        return tmp
\end{verbatim}

\subsection*{3| Tomonaga Luttinger Liquid (TLL)}
A TLL low-energy Hamiltonian in a box can be defined as\cite{Haldane_1981,PhysRevB.51.17827,PhysRevB.68.241301,PhysRevLett.79.5086,PhysRevX.9.011055,Zhu2022}
\[H_{TLL} = \frac{\pi\nu_{c}N^{2}}{4LK_{c}} + \frac{\pi\nu_{s}S^{2}_{s}}{LK_{s}}+\sum_{n=1}^{\infty}(\nu_{c}k_{n}a^{\dagger}_{c,n}a_{c,n}+\nu_{s}k_{n}a^{\dagger}_{s,n}a_{s,n}),\]

\noindent where $L$ is defect length, $c$ and $s$ label the charge and spin channels, $K_{c}$ and $K_{s}$ are two Luttinger parameters, $\nu_{c}$ and $\nu_{s}$ are charge and spin velocities, $N$ is the total electron filling, $S_{z}$ is the total z-component spin number, and $a^{\dagger}_{c,n}$ and $a^{\dagger}_{s,n}$ are the creation operators of charge and spin excitation. This representation identifies the key requirements for a TLL to be present within an ID. These requirements are such that 1) the first and second terms define a charging energy that determines the HOS-LUS energy gap ($E_{gap}$) arising from Coulomb interactions and the spin sector, respectively, 2) spin and charge show independent dispersions, 3) there exists $E_{gap}$ dependence that is inversely proportional to an ID length, and 4) creation operators give rise bosonic excitations of both spin and charge to yield a lateral and energetic dependence that exhibits particle-in-a-box behavior. In the results presented, all requirements are fulfilled to showcase the presence of a TLL hosted by an ID within WS$_2$.

\section*{Supplementary Figures}
\renewcommand{\figurename}{Supplementary Fig.}
\renewcommand{\thefigure}{1}
\begin{figure}[H]
    \centering
    \includegraphics[width = 1 \linewidth]{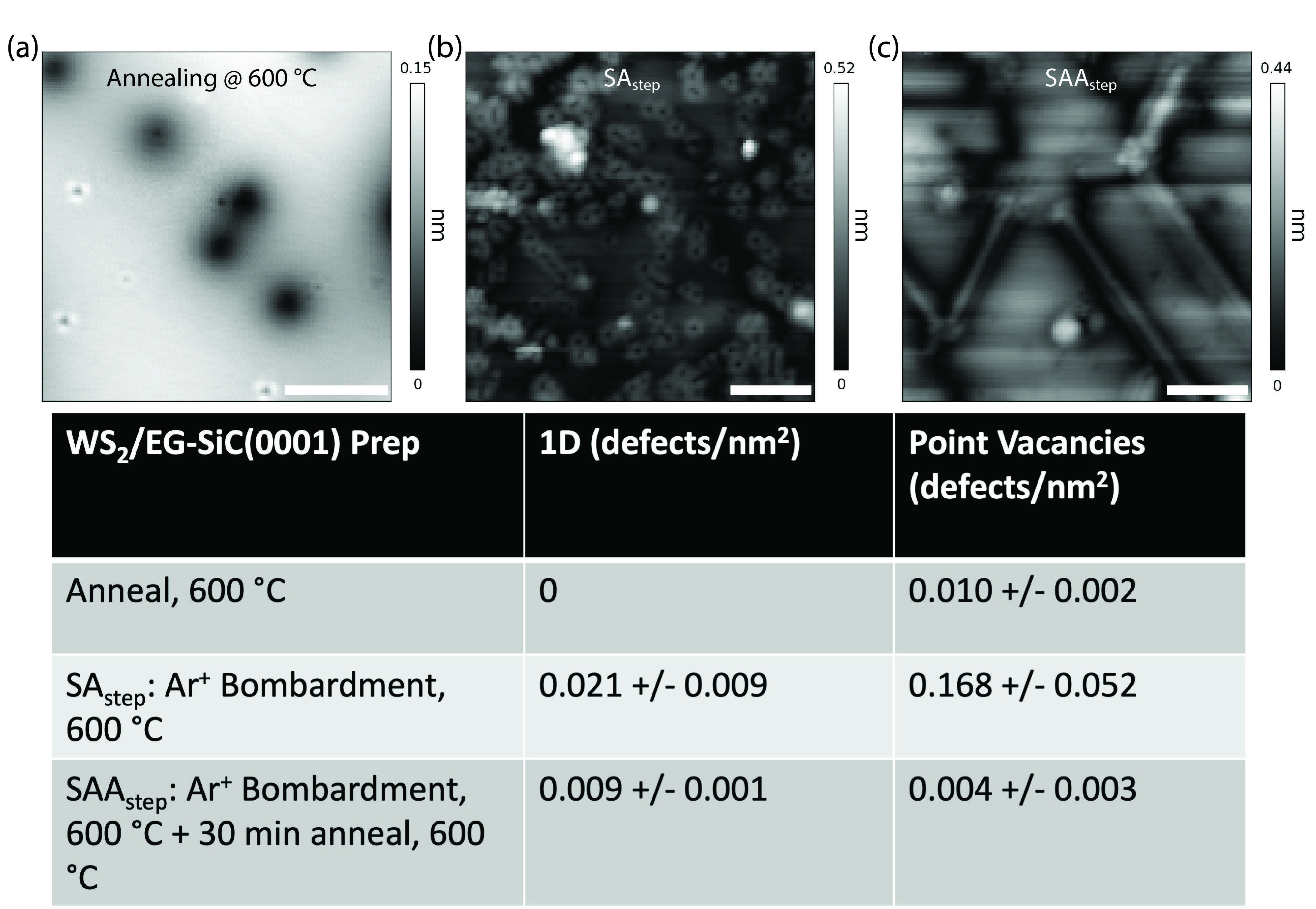}
    \caption{\textbf{Defect Density.} Defective density is calculated as the number of defects per unit area, where (a) annealed only samples show a low defect density, (b) sputtered with annealing produces a larger number of both 1D defects and V$_{\rm S}$, and (c) sputtered with annealing plus an additional 30 minute anneal produces elongated 1D defects with less V$_{\rm S}$ ($I_{tunnel}$ = 30 pA, $V_{sample}$ = 1.2 V). Scale bars, 4 nm. Each defect, across a large number of images over multiple samples and subsequent preparations, is selected by inspection and then by solving for the local minima or maxima within a given pixel window.}
    \label{figs1:fig_s1}
\end{figure}

\renewcommand{\thefigure}{2}
\begin{figure}[H]
    \centering
    \includegraphics[width = 1 \linewidth]{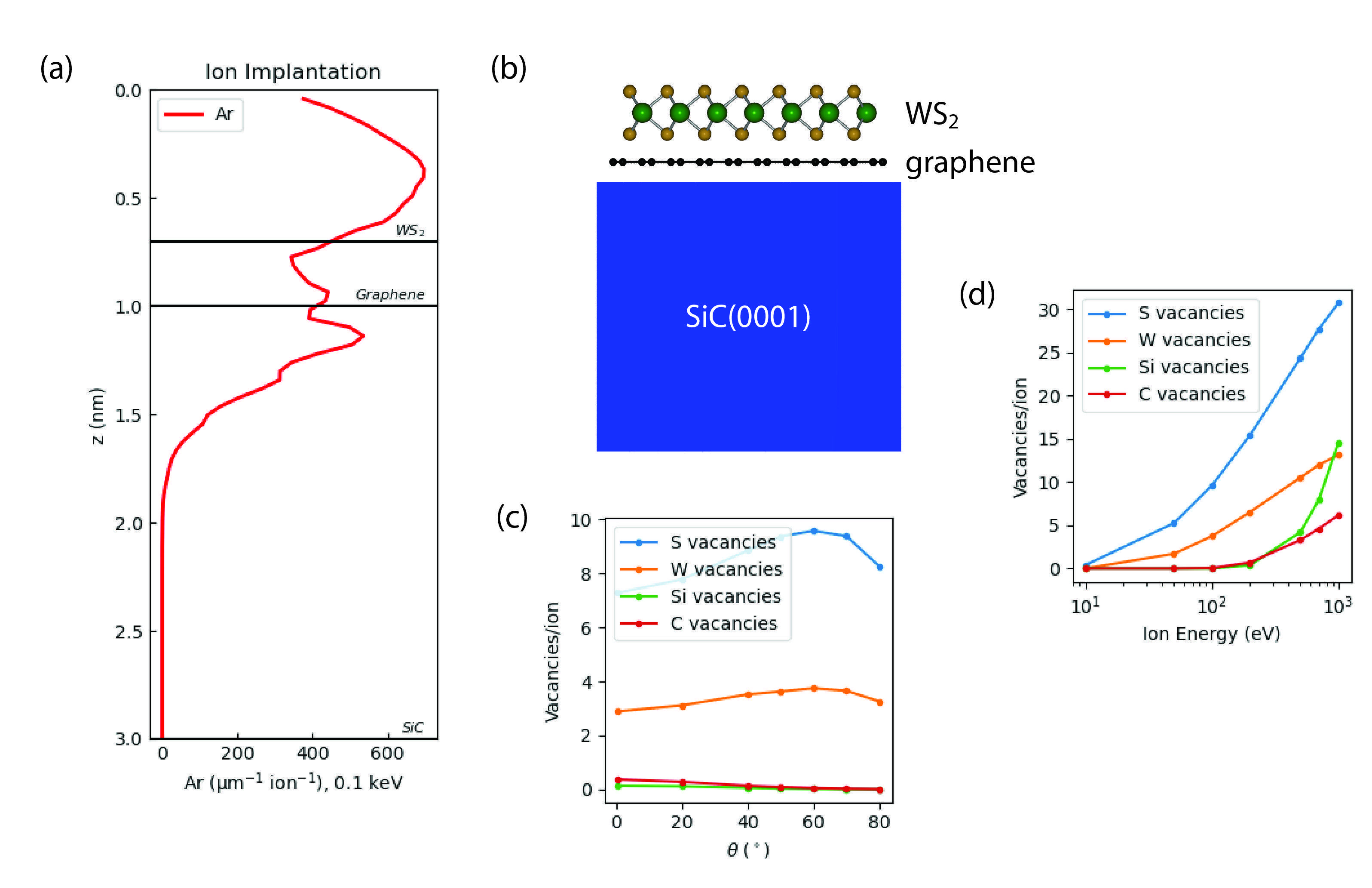}
    \caption{\textbf{SRIM Simulations.} (a) Results of SRIM simulations with 50000 ions for a (b) WS$_2$/Graphene/SiC(0001) heterostructure, where Ar$^+$ ions are expected to nominally interact with the TMD overlayer given the ion energy and angle of irradiation incidence. Both (c) and (d) depict the number of vacancies produced over a given incidence angle and energy, where we use an energy of 0.1 keV and an angle of 60$^{\circ}$, respectively.}
    \label{figs1:fig_s2}
\end{figure}

\renewcommand{\thefigure}{3}
\begin{figure}[H]
    \centering
    \includegraphics[width = 1 \linewidth]{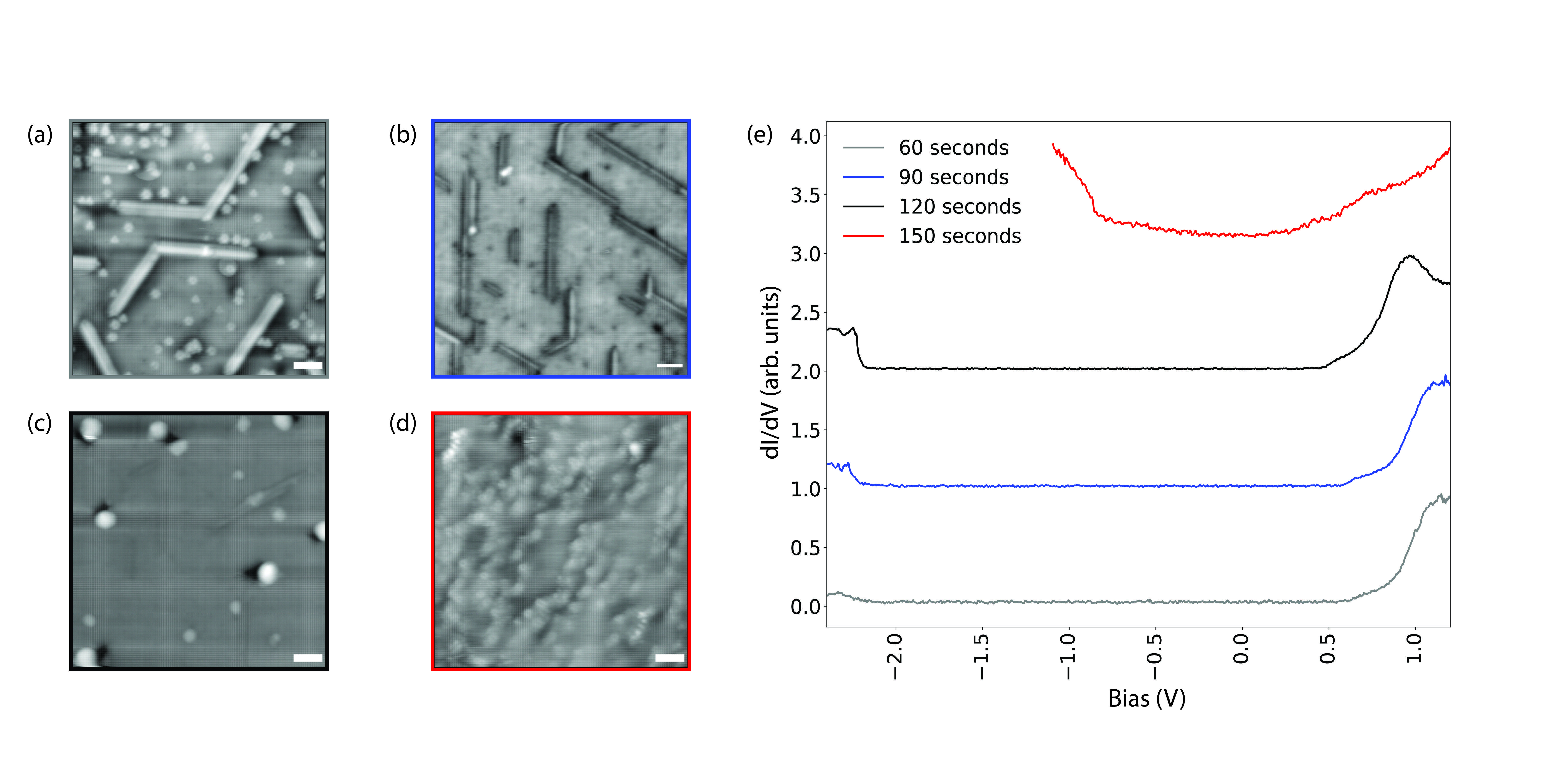}
    \caption{\textbf{1DM Elongation and WS$_2$ Degradation.} Constant current image over WS$_2$ after (a) two $SAA_{step}$ ($I_{tunnel}$ = 20 pA, $V_{sample}$ = 1.2 V), (b) three $SAA_{step}$ ($I_{tunnel}$ = 20 pA, $V_{sample}$ = 1.4 V), (c) four $SAA_{step}$ ($I_{tunnel}$ = 25 pA, $V_{sample}$ = 1.6 V), and (d) five $SAA_{step}$ ($I_{tunnel}$ = 25 pA, $V_{sample}$ = 1.6 V). Each cycle consists of a 30 second sputter and additional anneal. Scale bars, 4 nm. (e) dI/dV spectra recorded over non defective regions at each step and labeled by the total amount of sputter time ($V_{modulation}$ = 5 mV). The expected E$_{\rm gap}$ of WS$_2$ decreases above four $SAA_{step}$ and gains a metallic character, which is associated with material degradation.}
    \label{figs1:fig_s3}
\end{figure}

\renewcommand{\thefigure}{4}
\begin{figure}[H]
\vspace*{-55pt}
    \centering
    \includegraphics[width=0.8\linewidth]{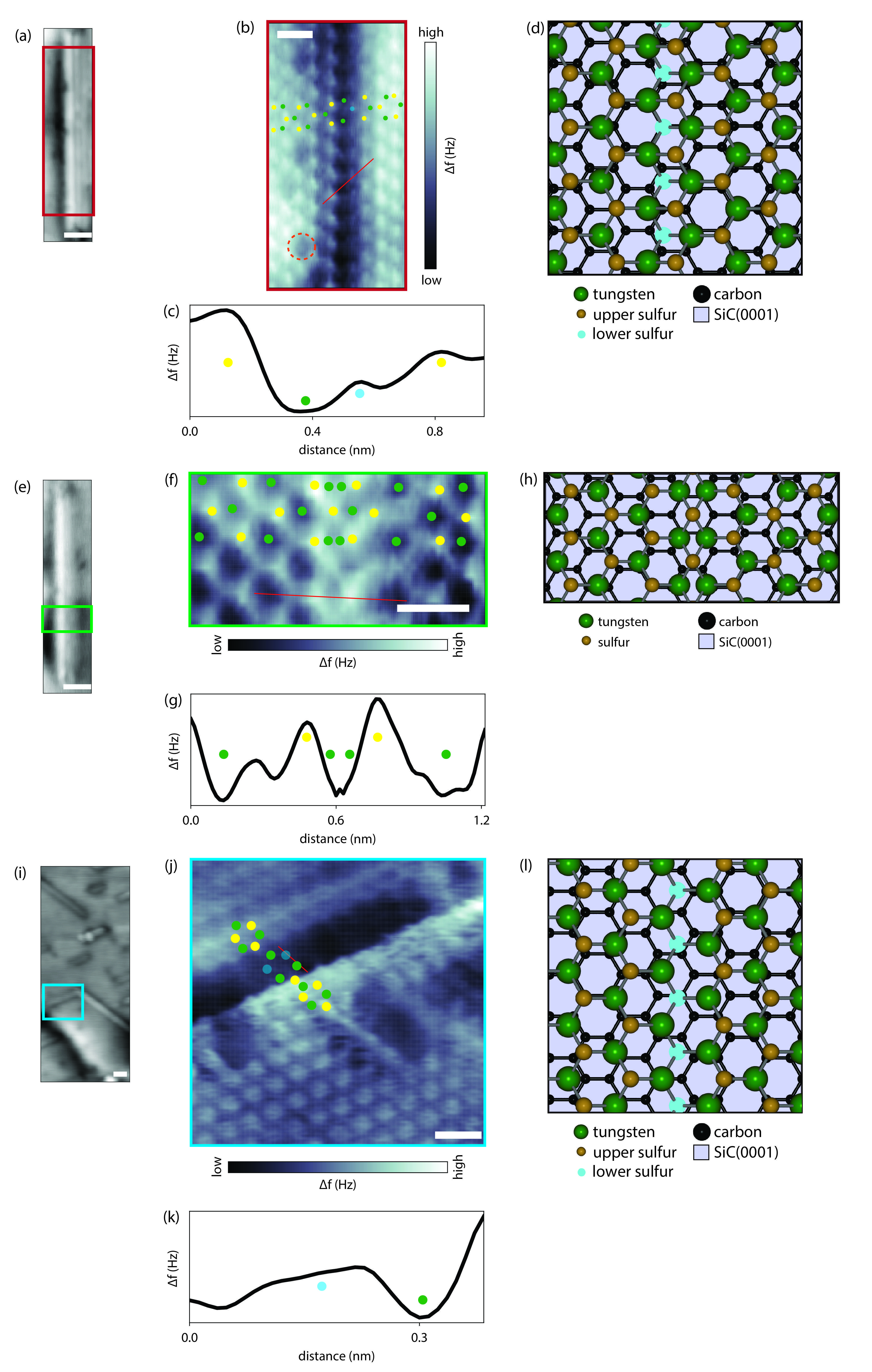}
    \caption{\textbf{Atomic Force Imaging.} (a) Constant current image over an isolated 4|4E intermediate 1DM defect ($I_{tunnel}$ = 30 pA, $V_{sample}$ = 1.2 V). Scale bar, 1.5 nm. (b) ncAFM ($V_{sample}$ = 0.0 V) collected over the same 4|4E intermediate 1DM with a ball model showing atomic locations. Scale bar, 0.4 nm. (c) A linescan over the red region shown in (b) that depicts two types of sulfur measured, where one is assigned to the top position (yellow) and the other to the bottom position (cyan) where metal sites appear as dips in frequency shift images. Depressions near the 4|4E intermediate 1DM reflect oxygen atom (circled in orange) chalcogen substitutions within an otherwise unmodified WS$_2$ lattice. A structure model is shown schematically in (d). A second isolated 1DM is shown in (e) that is identified to be an 4|4P intermediate 1DM ($I_{tunnel}$ = 30 pA, $V_{sample}$ = 1.2 V). Scale bar, 1.5 nm. (f) ncAFM confirms the location of top sulfur, tungsten, and hollow sites ($V_{sample}$ = 0.0 V), where a linescan of across the 4|4P intermediate 1DM in (g) shows the location of dual tungsten surrounded by sulfur sites that is also shown schematically in (h). Scale bar, 0.4 nm. A 1DM chalcogen vacancy line is shown in (i) ($I_{tunnel}$ = 30 pA, $V_{sample}$ = 1.2 V). Scale bar, 1.5 nm. (j) Sulfur vacancies are measured in ncAFM ($V_{sample}$ = 0.0 V). A linescan across the 1DM in (k) shows the tungsten and lower sulfur sites, schematically depicted in (l). Scale bar, 0.4 nm.}
    \label{figs1:fig_s3}
\end{figure}

\renewcommand{\thefigure}{5}
\begin{figure}[H]
    \centering
    \includegraphics[width = 0.8 \linewidth]{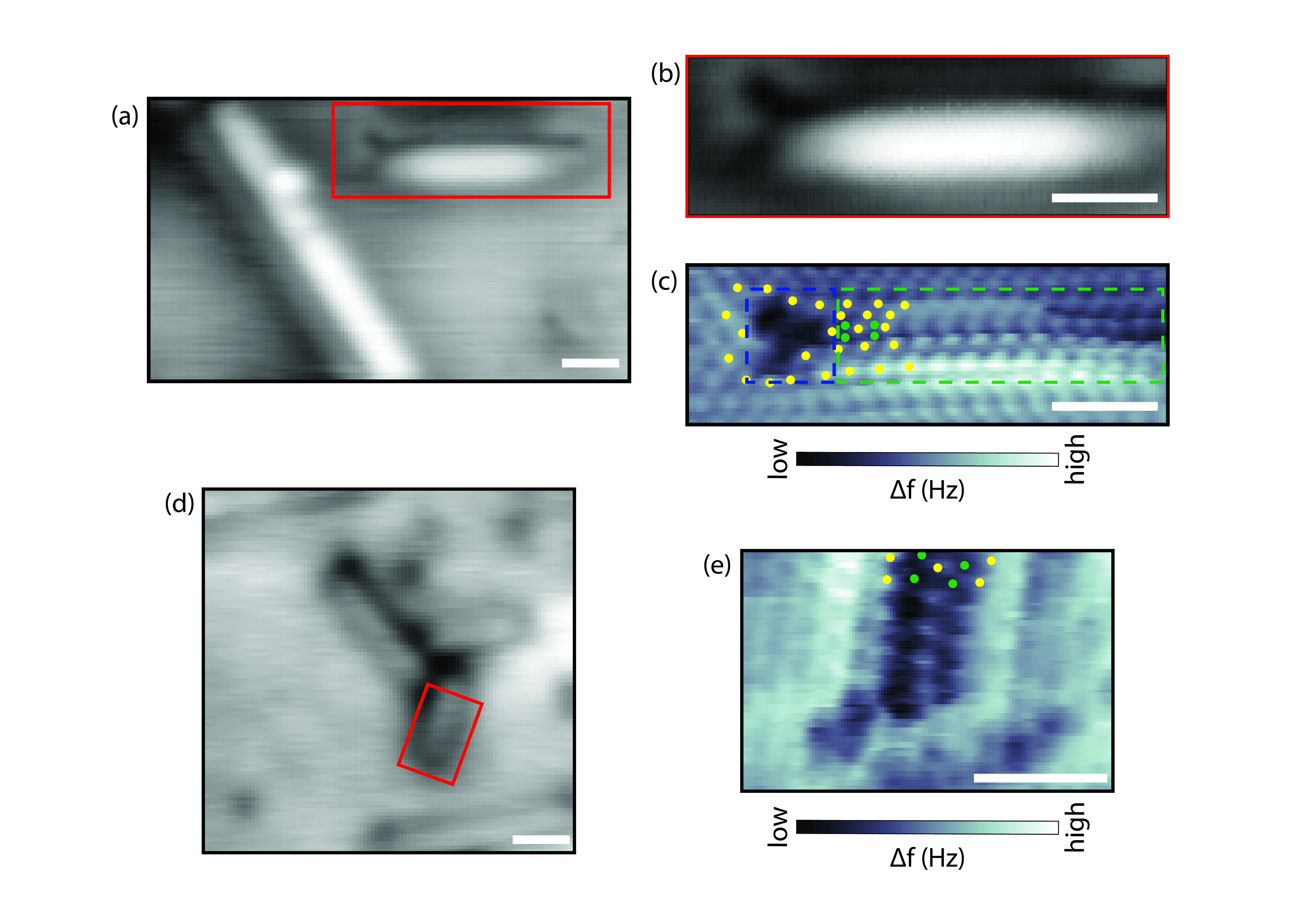}
    \caption{\textbf{1DM Defect End Point Mapping.} (a) Constant current image over an isolated 1DM defect ($I_{tunnel}$ = 30 pA, $V_{sample}$ = 1.2 V) terminating in an end point defect. Scale bar, 1 nm. (b) Zoomed in image that is highlighted in (a) with a red box, where sequential (c) ncAFM with a CO functionalized tip ($V_{sample}$ = 0.0 V) collected over the same 1DM showcase local structure. Scale bars, 1 nm. A chalcogen depletion region (blue) is located at the edge of the isolated and strained 1DM(further highlighted in green). Chalcogen (sulfur) sites are highlighted with yellow spheres and metal (tungsten) sites are highlighted with green spheres. (d) Another defect is measured that appears as an intermediate formation before a fully relaxed MTB ($I_{tunnel}$ = 30 pA, $V_{sample}$ = 1.2 V). Scale bar, 1nm. ncAFM over an endpoint region (highlighted with a red box) shows strained metallic-rich regions that form into a 4|4P intermediate structure.}
    \label{figs1:fig_s4}
\end{figure}

\renewcommand{\thefigure}{6}
\begin{figure}[H]
    \centering
    \includegraphics[width = 1.0 \linewidth]{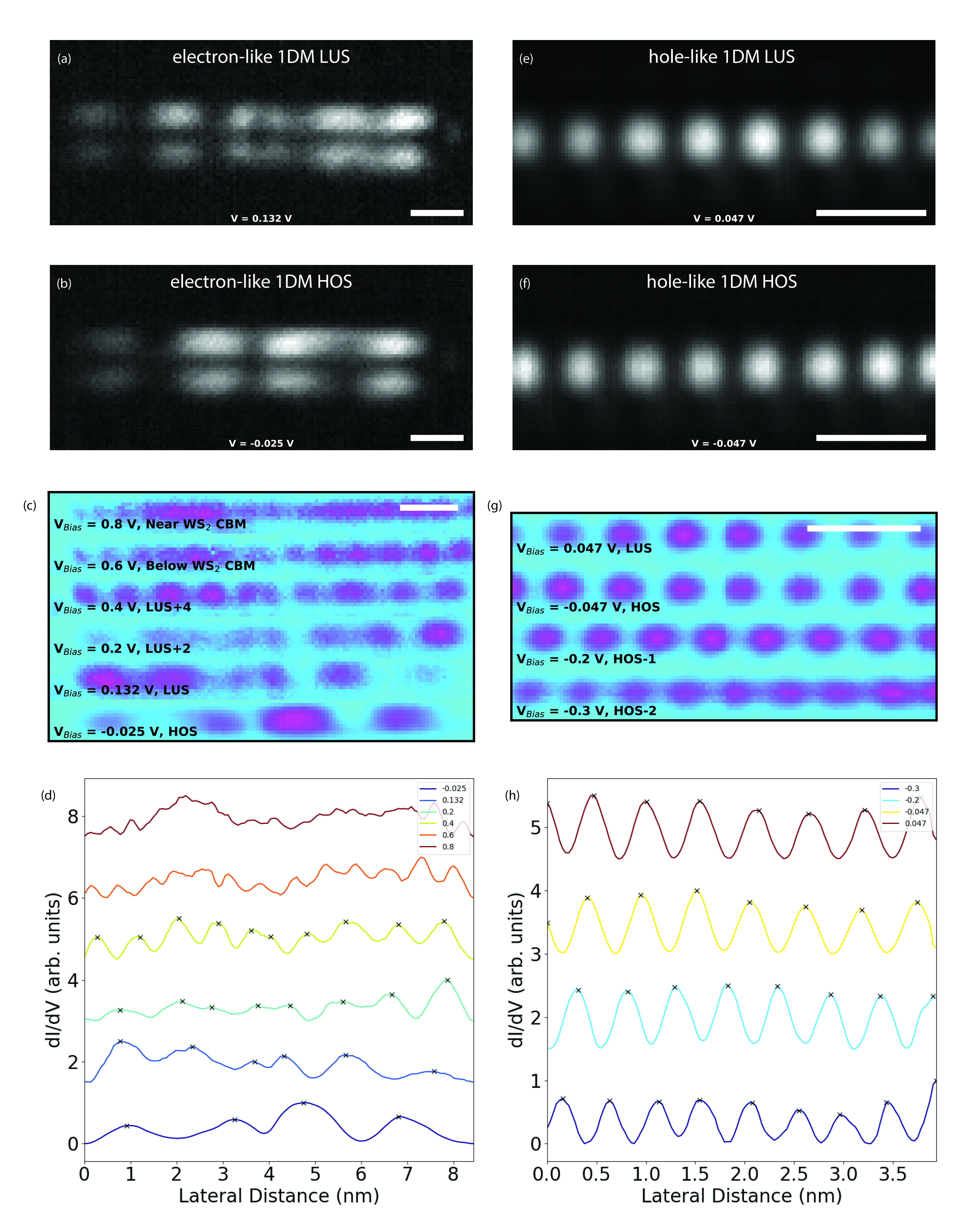}
    \caption{\textbf{LDOS Mapping.} Conductance maps ($V_{modulation}$ = 5 mV) performed across an electron-like 1DM show dual-line orbital behavior at (a) 0.132 eV (LUS) and (b) -0.025 eV (HOS) that is spatially out of phase. (c) Accumulated conductance maps across a single-line of the electron-like 1DM ($V_{modulation}$ = 5 mV) are further shown as a function of bias, where the number of nodes increase as bias voltage is increased. Scale bars, 1 nm. Peak assignments can be made by solving for local maxima along a line profile, which is compiled in (d). Conductance maps (dI/dV) of the as-measured (e) 0.047 eV (LUS) and (f) -0.047 eV (HOS) that are spatially in phase within an hole-like 1DM. (g) Compiled conductance maps ($V_{modulation}$ = 5 mV) across the single-line hole-like 1DM are further shown, where the number of nodes decreases as the voltage is increased. (h) Line profiles extracted from (g) detailing local maxima.}
    \label{fig2:spec}
\end{figure}

\renewcommand{\thefigure}{7}
\begin{figure}[H]
    \centering
    \includegraphics[width = 1 \linewidth]{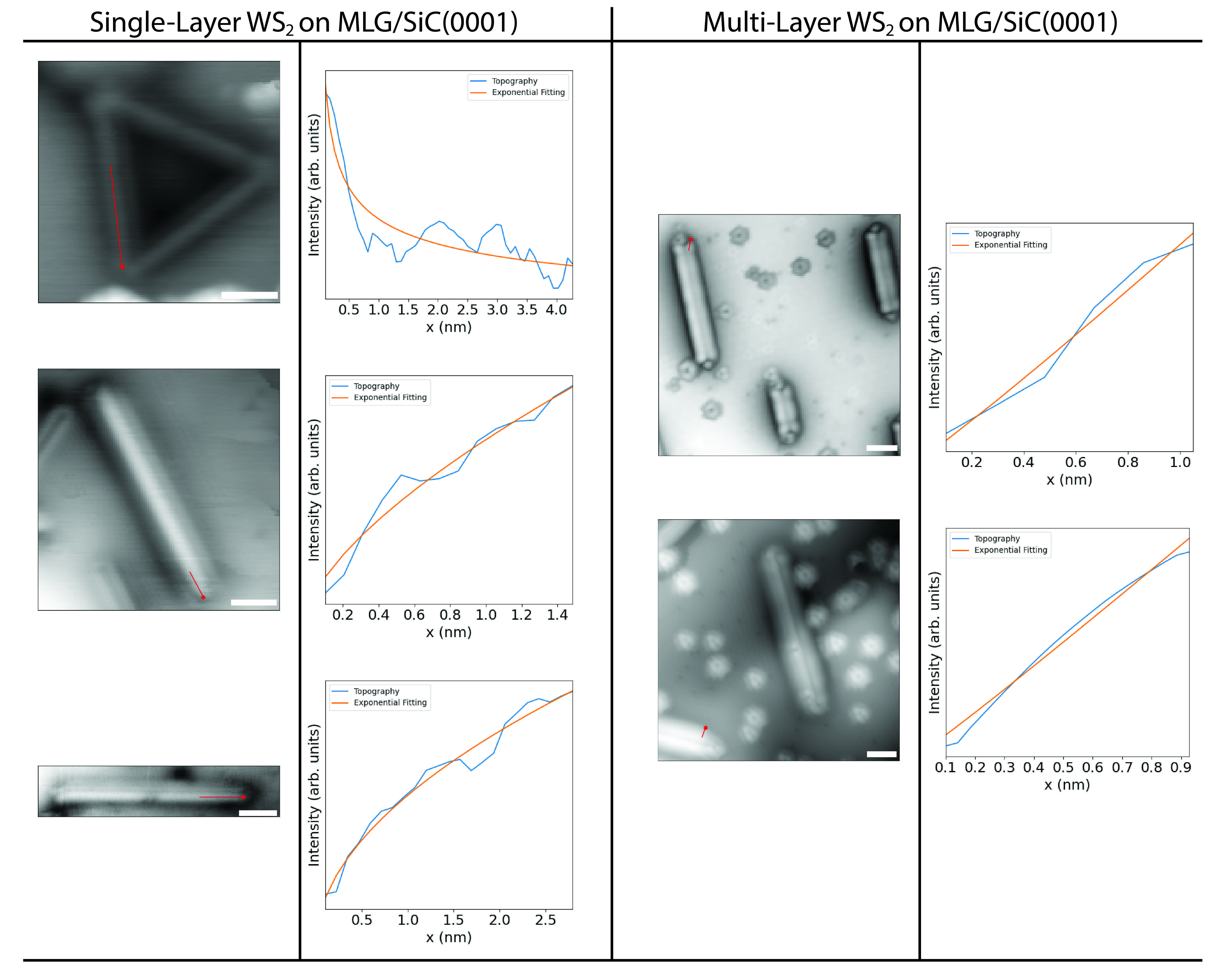}
    \caption{\textbf{Constant-Current Density of States Decay Across a 1DM.} A power-law dependence is measured across a multiple 1DM defects on both monolayer WS$_2$ and multilayer (3ML) WS$_2$ on MLG/SiC(0001). In each image depicted ($V_{sample}$ = 1.2 V), multiple fits were taken across defects shown, where representative linescans beginning at a starred point and along a given line (red) are shown with corresponding constant-current profiles with subsequent exponential fittings. Across 14 lineprofiles and 10 defects, multilayer WS$_2$ defects show an absolute power-law exponential parameter of 0.99 $\pm$ 0.15, which indicates Fermionic behavior, and monolayer WS$_2$ defects yield a parameter of 0.49 $\pm$ 0.16 that matches expected behavior of a Luttinger liquid. Scale bars, 2nm. Fittings were performed using the lmfit package in Python\cite{newville_matthew_2014_11813}.}
    \label{figs1:fig_s9}
\end{figure}

\renewcommand{\thefigure}{8}
\begin{figure}[H]
    \centering
    \includegraphics[width = 1.0 \linewidth]{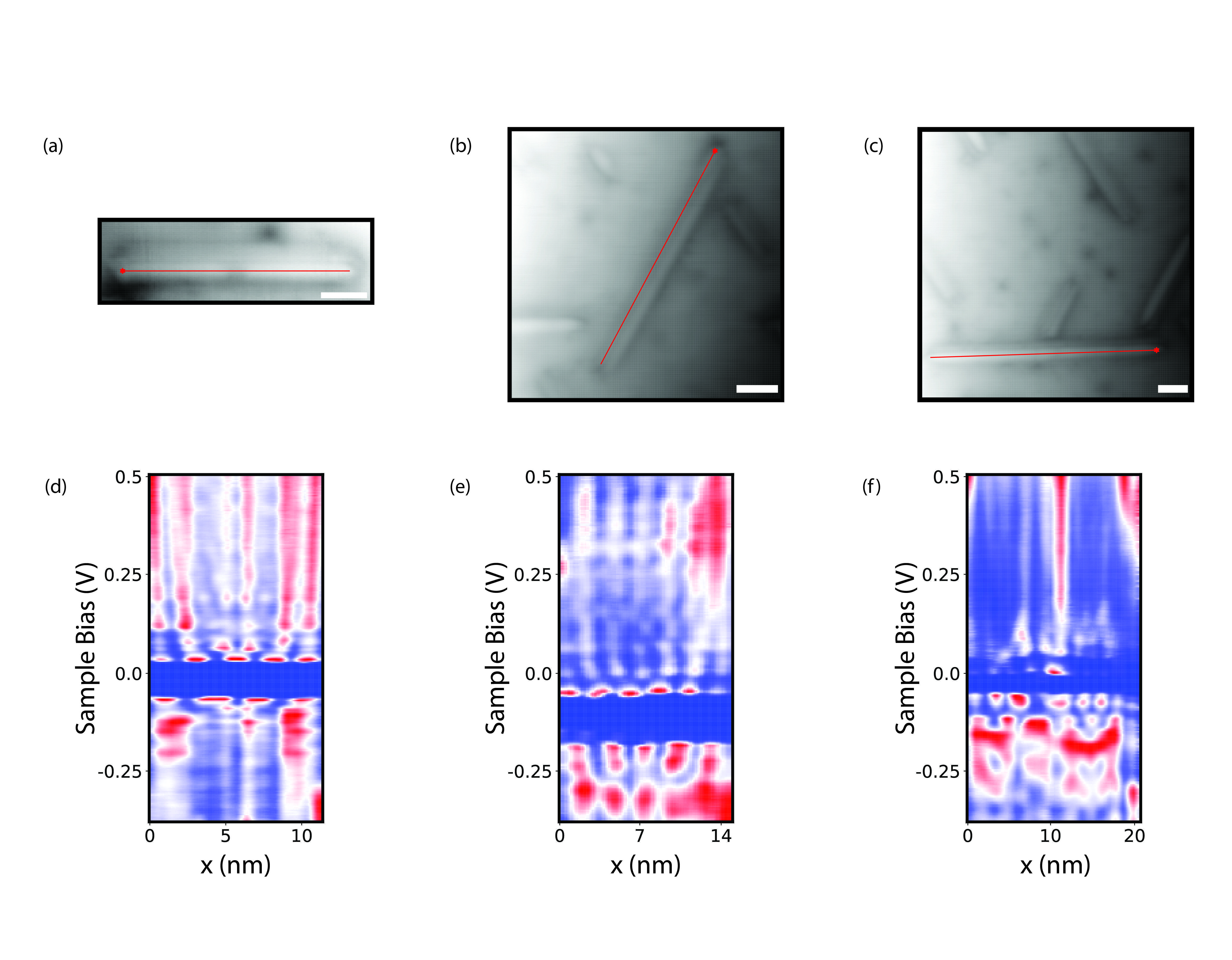}
    \caption{\textbf{Electron-like 1DM Dispersion Length Dependence.} Electron-like dispersions are collected on 1DMs with different lengths. Constant current images over 1DMs that are (a) 11.19 nm ($I_{tunnel}$ = 30 pA, $V_{sample}$ = 1.2 V), (b) 14.86 nm ($I_{tunnel}$ = 30 pA, $V_{sample}$ = 1.4 V), and (c) 20.44 nm ($I_{tunnel}$ = 80 pA, $V_{sample}$ = 1.2 V) are collected. Scale bars, 2 nm. Dense scanning tunneling spectra ($V_{modulation}$ = 5 mV, $I_{set}$ = 150 pA) are collected along the red line, beginning at the starred point. Results are shown for the (d) 11.19 nm defect (1x128x500 pixels), (e) 14.86 nm defect (1x128x400 pixels), and the (f) 20.44 nm defect (1x128x400 pixels). Using the HOS as a reference, a node spacing of is measured.}
    \label{fig2:spec}
\end{figure}

\renewcommand{\thefigure}{9}
\begin{figure}[H]
    \centering
    \includegraphics[width = 1 \linewidth]{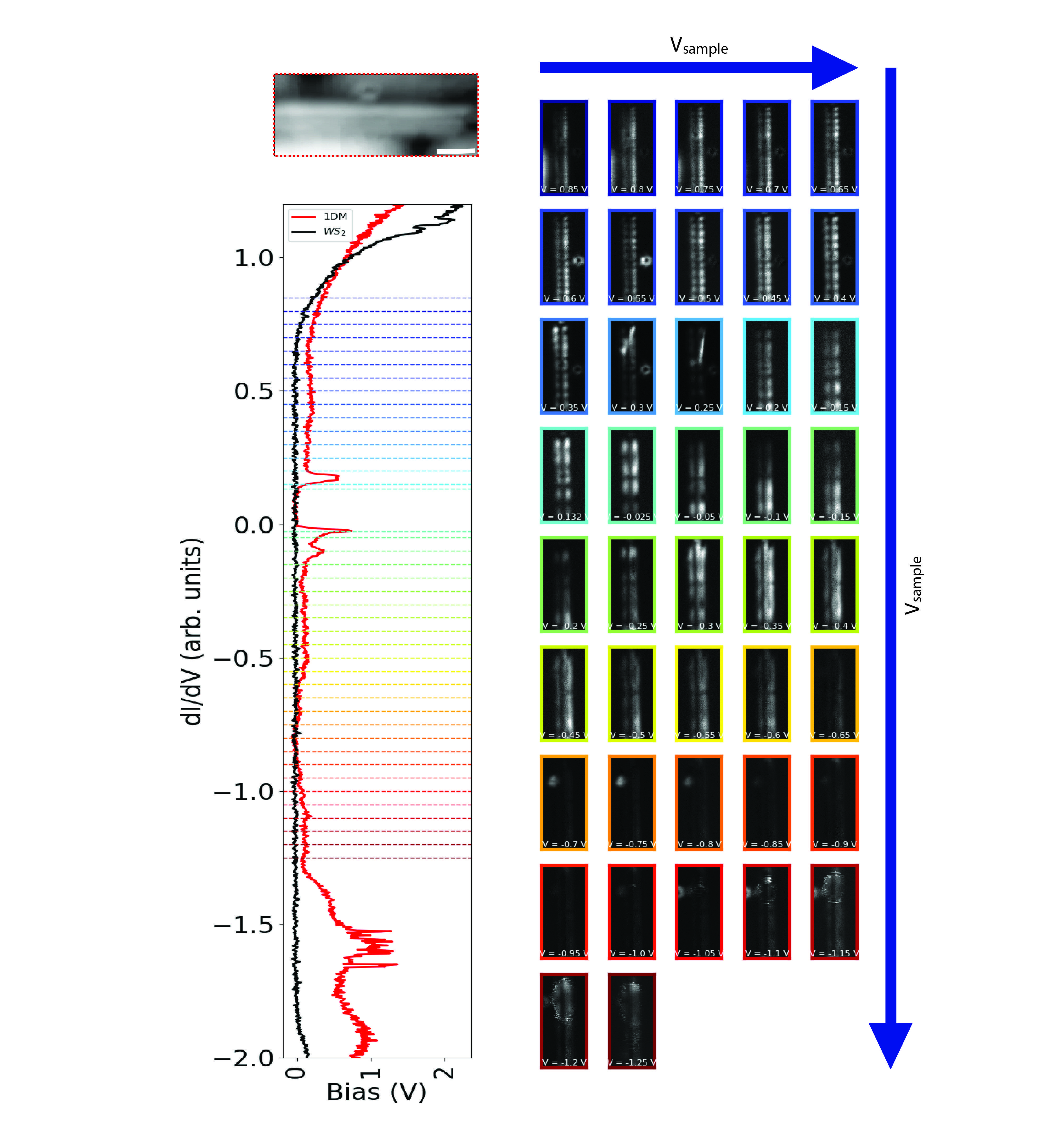}
    \caption{\textbf{Differential Conductance Mapping.} dI/dV mapping ($V_{modulation}$ = 5 mV) over the spectra region shown for a electron-like 1DM, with 4|4P intermediate structure, on a jet color scale, where the energy is ramped from near the VBM of WS$_2$ by 0.05 V to the HOS gap opening of the 1DM hosting a TLL, and then from the LUS to the CBM of WS$_2$. Arrows indicate decreasing bias. The defect imaged is shown to the upper left, where dI/dV images are representative of the region highlighted in dashed red ($I_{tunnel}$ = 30 pA, $V_{sample}$ = 1.2 V). Scale bar, 1.5 nm. A 1D particle in a box behavior is evident, and orbitals of both the TLL and a V$_{\rm S}$ can be visualized on as-acquired data at respective energies. Additionally, presence of a V$_{\rm S}$ scatters available quantum-well states (spatially-centered within the defect) above the HOS.}
    \label{figs1:fig_s5}
\end{figure}

\renewcommand{\thefigure}{10}
\begin{figure}[H]
    \centering
    \includegraphics[width = 1 \linewidth]{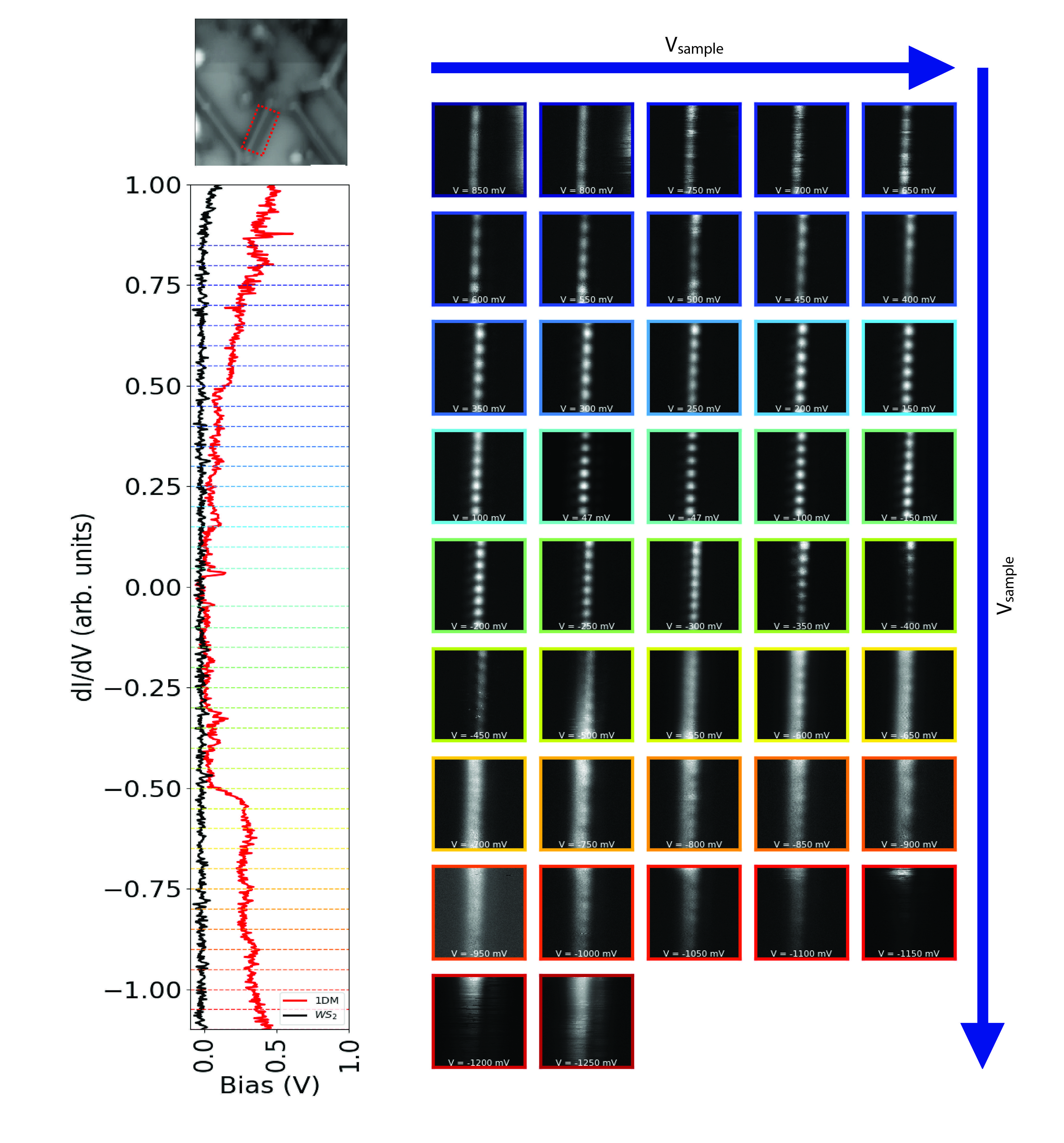}
    \caption{\textbf{Differential Conductance Mapping.} dI/dV mapping ($V_{modulation}$ = 5 mV) over the spectra region shown for a hole-like 1DM, with 4|4E intermediate structure, on a jet color scale, where the energy is ramped from near the VBM of WS$_2$ by 0.05 V to the HOS, and then from the LUS to the CBM of WS$_2$. The defect imaged is shown to the upper left, where the partial portion of the mapped defect is highlighted in dashed red ($I_{tunnel}$ = 30 pA, $V_{sample}$ = 1.2 V). Scale bar, 4 nm. Orbitals of the as-formed TLL can be visualized on as-acquired data as a function of bias voltage, where arrows indicate decreasing bias.}
    \label{figs1:fig_s6}
\end{figure}

\renewcommand{\thefigure}{11}
\begin{figure}[H]
    \centering
    \includegraphics[width = 0.8 \linewidth]{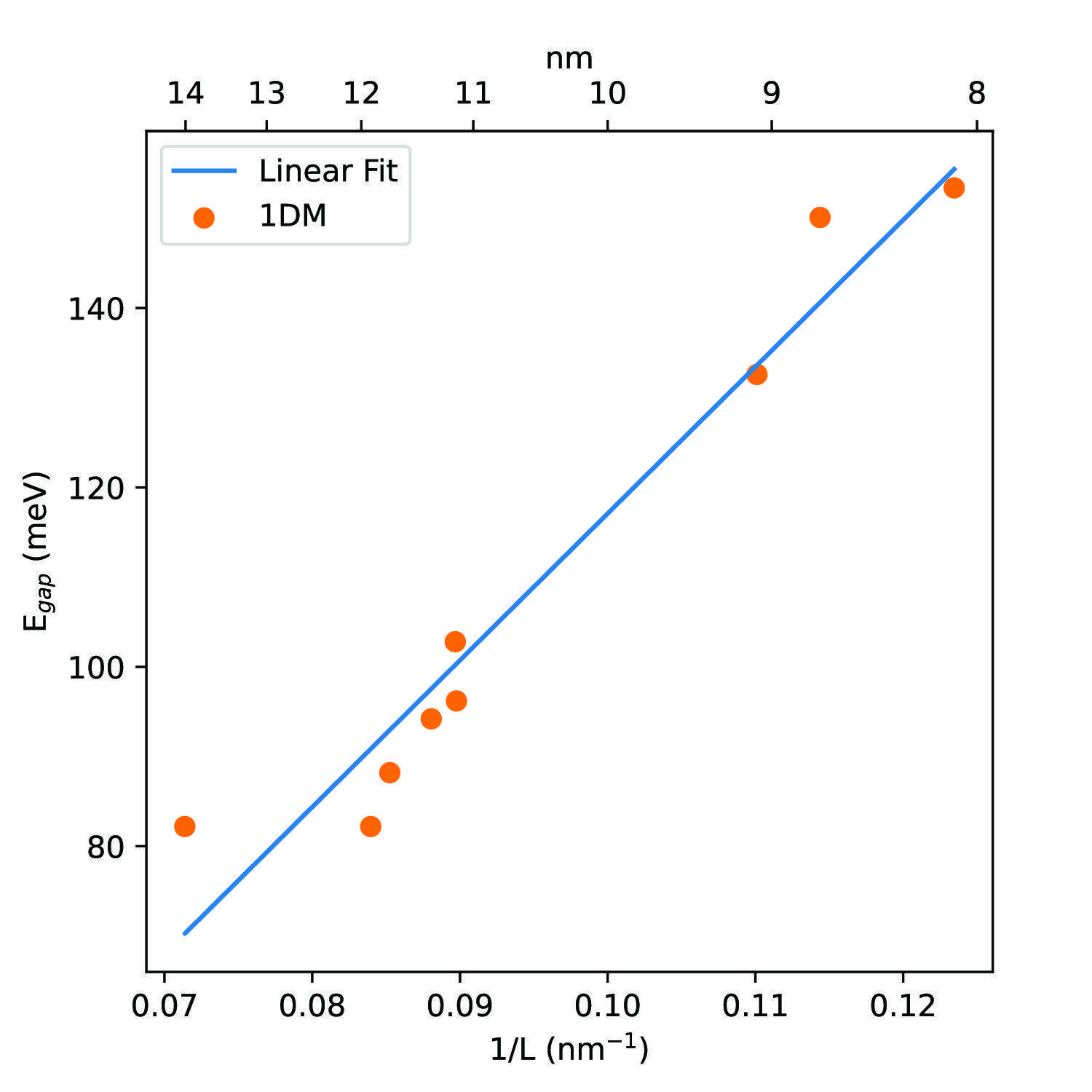}
    \caption{\textbf{Gap Length Dependence.} Band gaps are depicted across 9 1DMs, where each point represents the average of multiple reproducible data points, measured as a function of length (slope = 1636.4 $\pm$ 152.0 meV$\cdot$nm, offset = -46.5 $\pm$ 14.7 meV). A linear relationship is shown across both 1DM structures. Fitting was performed using the lmfit package in Python\cite{newville_matthew_2014_11813}.}
    \label{figs1:fig_s8}
\end{figure}

\renewcommand{\thefigure}{12}
\begin{figure}[H]
    \centering
    \includegraphics[width = 0.6 \linewidth]{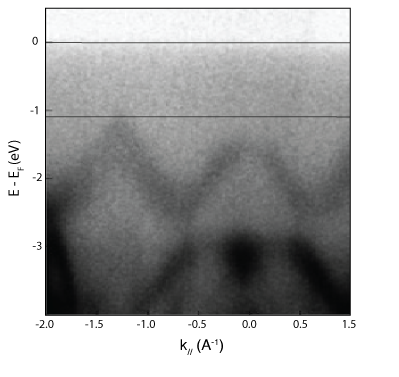}
    \caption{\textbf{nARPES Polarization Investigation.} WS$_2$ bands collected with linear vertical polarization. The horizontal line below E-E$_F$ indicate the top of the valence band for the defective crystal.}
    \label{figs1:fig_s10}
\end{figure}

\renewcommand{\thefigure}{13}
\begin{figure}[H]
    \centering
    \includegraphics[width = 0.6 \linewidth]{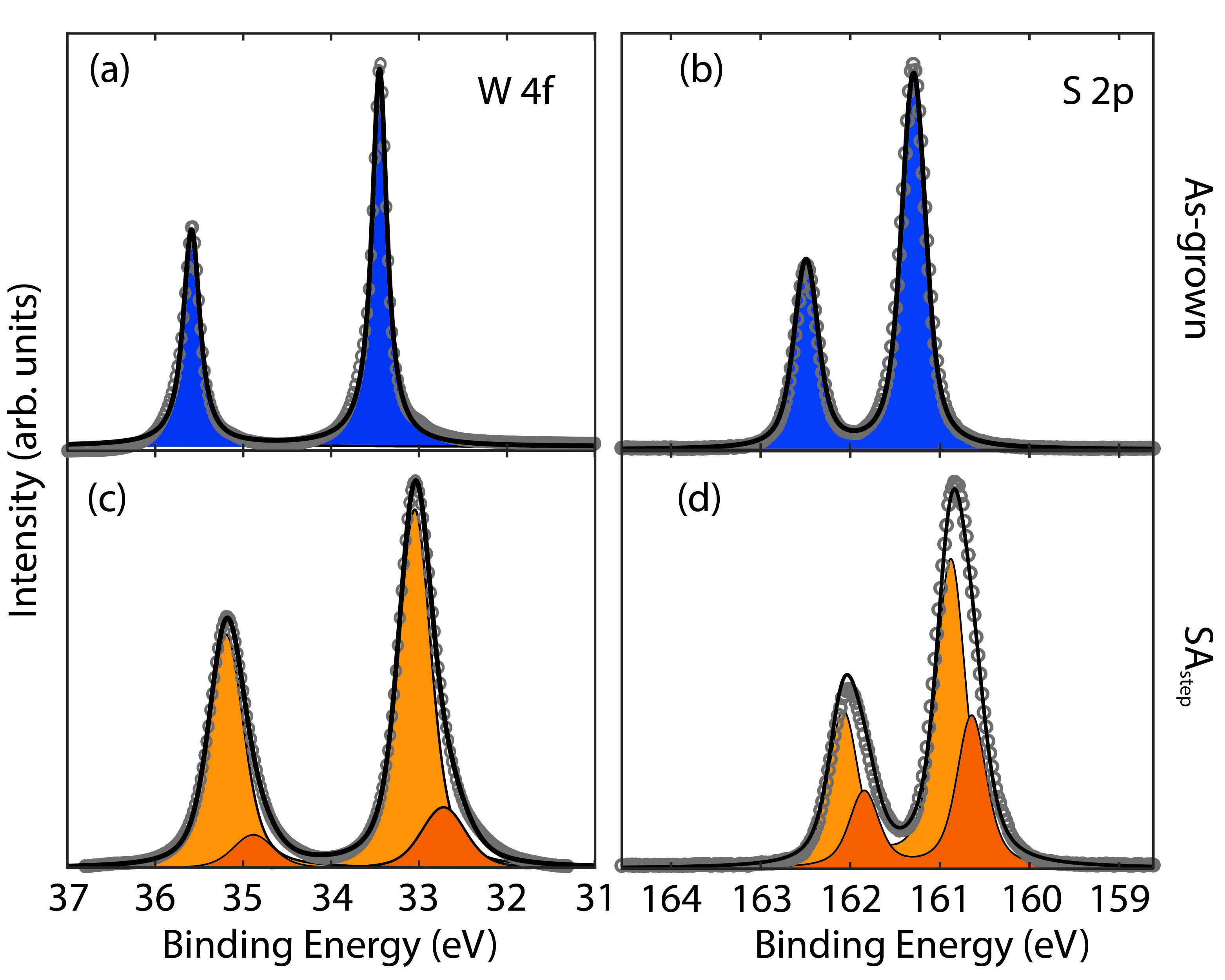}
    \caption{\textbf{Measured W and S Core Levels Spectra of As-grown and Defective WS$_2$.} (a) W 4$f$ 7/2 and 5/2 levels centered at 33.5 eV and 35.7 eV, respectively, and (b) S 2$p$ 3/2 and 1/2 core levels centered at 161.3 eV and 162.4 eV, respectively, from the unmodified sample. (c) and (d) are relative to WS$_2$ after $SA_{step}$, displaying two components (light and dark orange), with a relative shift of 0.4 eV for W 4$f$ peaks and 0.2 eV for S 2$p$ peaks.}
    \label{figs1:fig_s11}
\end{figure}

\renewcommand{\thefigure}{14}
\begin{figure}[H]
    \centering
    \includegraphics[width = 1.0 \linewidth]{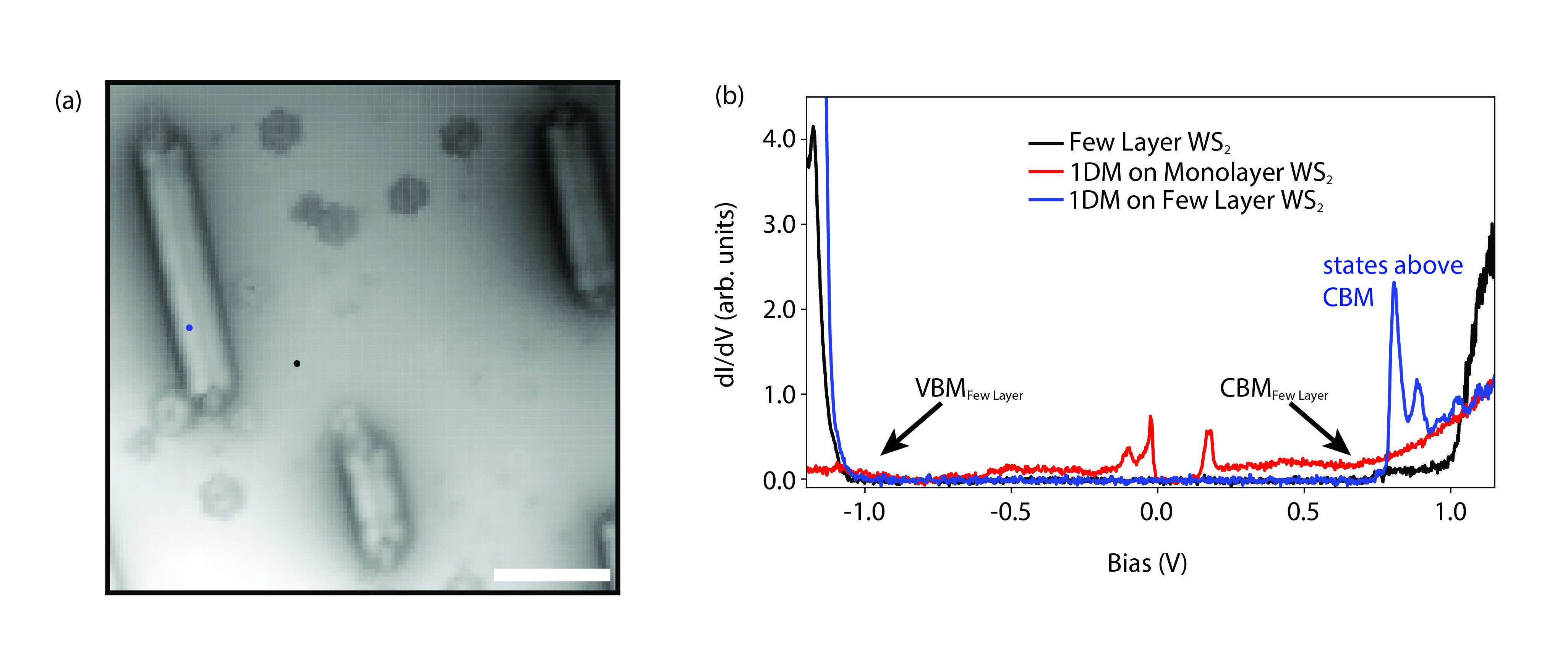}
    \caption{\textbf{LDOS on Multilayered WS$_2$.} (a) Scanning tunneling micrograph depicting 3 ML of WS$_2$ over MLG/SiC(0001) with both IDs and point defects ($I_{tunnel}$ = 30 pA, $V_{sample}$ = 1.2 V). Scale bar, 4 nm. (b) dI/dV point spectroscopy of 3 ML WS$_2$ (black), a 1DM within 1 ML WS$_2$ (red), and a 1DM within 3 ML WS$_2$ (blue). Corresponding spectral locations are depicted in (a), where both the VBM onset and CBM onset for the ID (blue) match that of the as-grown 3 ML WS$_2$ (black) spectra.}
    \label{fig5:multilayer}
\end{figure}

\renewcommand{\thefigure}{15}
\begin{figure}[H]
    \centering
    \includegraphics[width = 0.6 \linewidth]{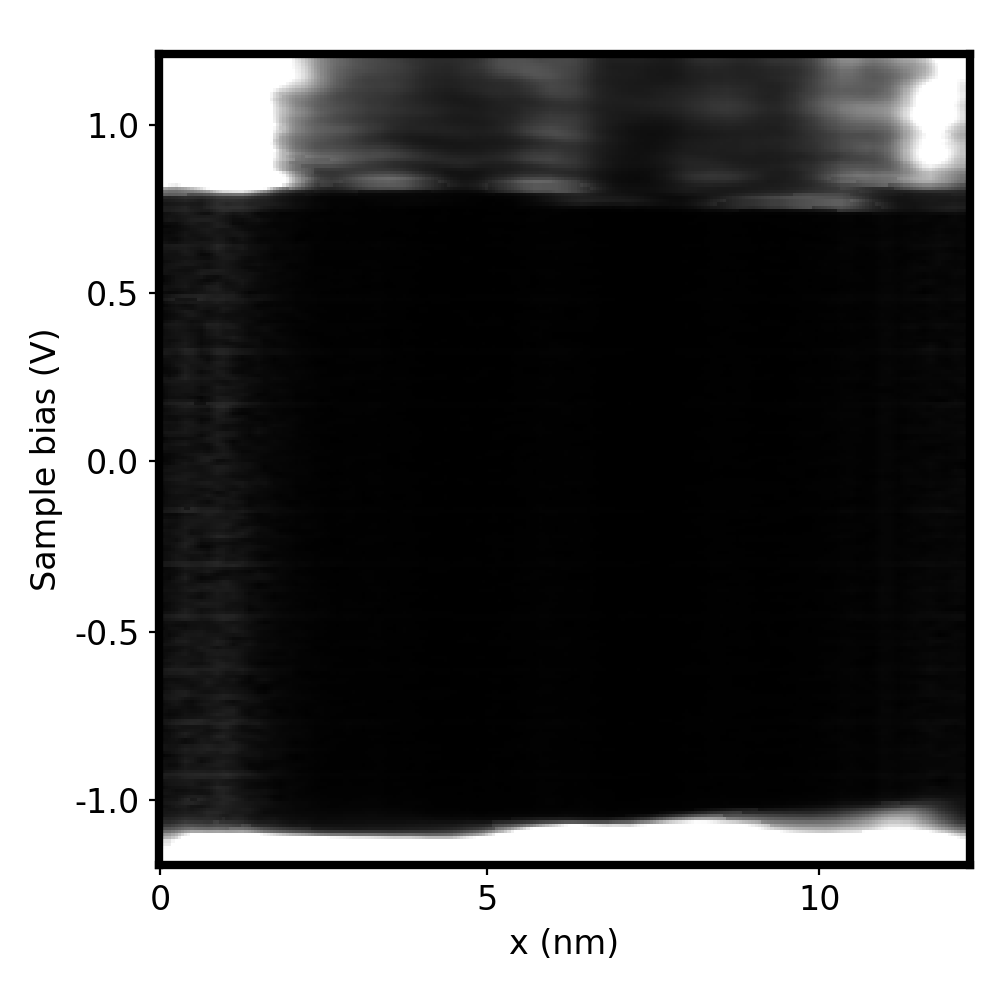}
    \caption{\textbf{Spatially Resolved Scanning Tunneling Spectroscopy over Multilayer WS$_2$.} Dense LDOS spectra (1x128x500 pixels) collected over an anticipated 1DM not in contact with underlying graphene ($V_{modulation}$ = 5 mV, $I_{set}$ = 150 pA), where states within the gap are not present.}
    \label{figs1:fig_s7}
\end{figure}
